\documentclass[12pt]{article}
\pdfoutput=1

\usepackage{paper2e}
\usepackage{mydefs2e}
\usepackage{xspace}
\usepackage{graphicx}
\usepackage{amssymb}

\newcommand{\Sla}[1]%
{\kern0.12em{\raise.15ex\hbox{$/$}\kern-.74em #1}}

\newcommand{\MS}{\ensuremath{M_{\rm SUSY}}\xspace}

\newcommand{\lag}{\mathcal L}
\renewcommand{\d}{\partial}

\newcommand{\beqa}{\begin{eqnarray}}
\newcommand{\eeqa}{\end{eqnarray}}

\begin{document}

\begin{titlepage}

\title{Superconformal Technicolor:\\\medskip
Models and Phenomenology}

\author{Aleksandr Azatov$^*$,\ \ 
Jamison Galloway$^*$,\ \ 
Markus A. Luty$^\dagger$} 

\address{$^*$Dipartimento di Fisica, Universit\`a di Roma ``La Sapienza"\\
and INFN Sezione di Roma, I-00185 Roma}

\address{$^\dagger$Physics Department, University of California Davis\\
Davis, California 95616}

\begin{abstract}
In supersymmetric theories with a strong conformal sector,
soft supersymmetry breaking naturally gives
rise to confinement and chiral symmetry breaking 
in the strong sector at the TeV scale.
We construct and analyze models where such a sector dynamically
breaks electroweak symmetry, and take the first steps in
studying their phenomenology.
We consider two scenarios, one where the strong dynamics induces
vacuum expectation values for elementary Higgs fields,
and another where the strong dynamics is solely responsible
for electroweak symmetry breaking. 
In both cases there is no fine tuning required to explain
the absence of a Higgs boson below the LEP bound,
solving the supersymmetry naturalness problem.
Quark and lepton masses arise from conventional
Yukawa couplings to elementary Higgs bosons, so there
are no additional flavor-changing effects associated with the strong
dynamics.
A good precision electroweak fit can be obtained
because the strong sector is an $SU(2)$ gauge theory
with one weak doublet, and has adjustable parameters that
control the violation of custodial symmetry.
In addition to the the  standard supersymmetry signals, 
these models predict production of multiple heavy standard
model particles ($t$, $W$, $Z$, and $b$) from decays of
resonances in the strong sector.
The strong sector has no approximate parity symmetry,
so $WW$ scattering is unitarized by states that can
decay to $WWW$ as well as $WW$.
\end{abstract}

\end{titlepage}

\section{Introduction}
\label{sec:intro}
Supersymmetry (SUSY) gives a compelling solution to the electroweak
hierarchy problem, and provides a sensible framework for speculations
about physics above the TeV scale.
It is for this reason that so much of the theoretical and experimental
effort in physics beyond the standard model is devoted to SUSY.
However, if SUSY is the solution of the hierarchy problem it generically
predicts a standard-model-like
Higgs boson with mass below $m_Z$, which is ruled out.
In the MSSM, this can be avoided only by radiative corrections that
introduce fine tuning at the percent level.
It is possible to avoid this tuning by extending the MSSM, either to
raise the Higgs mass \cite{raisetheHiggs}
or to give it new decays that are less constrained
by experiment \cite{newdecays},
but the models must be carefully constructed to have these features.

Technicolor also gives a compelling solution to the hierarchy problem,
but it is generally considered less plausible than SUSY mainly because
of problems with flavor and precision electroweak tests.
The traditional approach to incorporating flavor into technicolor
theories involves extending the gauge group of technicolor 
to include the flavor symmetries, which are then broken above
the TeV scale \cite{TCreview}.
There are daunting obstacles to constructing realistic models of 
this kind, and there is no realistic example in the literature.
Furthermore, any such model must have large numbers of
technicolors and/or techniflavors, and therefore is expected to give
large corrections to the precision electroweak
parameters $S$ and $T$ that are incompatible with data.
The prospects are much better if the couplings responsible
for quark and lepton masses arise from the exchange of heavy scalars
\cite{TCscalar}.
This is potentially natural in supersymmetric models, where SUSY is
broken above the TeV scale.
In this case, the higher-dimension operators that generate quark
and lepton masses can be generated from exchange of Higgs scalars,
which can incorporate minimal flavor violation and do not require
extending the technicolor gauge sector.
The pioneering attempts in this direction \cite{bosonicTC}
cannot accommodate the large observed value
of the top quark mass, but realistic models
have recently been constructed \cite{CTCflavor} in the
context of conformal technicolor \cite{CTC}.
These are explicit UV complete models with a minimal technicolor
sector at a TeV, that do not conflict with precision
electroweak and flavor constraints.

In this paper, we combine SUSY and conformal technicolor in a more
direct way in an attempt to address the shortcomings of both.
(For recent closely related work, see \Ref{Kagan}.)
A companion paper \cite{SCTCshort} describes the main ideas and results
in a succinct fashion, while this paper gives a full discussion.
This paper is written to be self-contained, and can be read
on its own.
%

We assume that the visible sector consists of the MSSM plus a
strong sector.
SUSY is assumed to be broken at the TeV scale in both the
MSSM and the strong sector, as is natural in many theories of
SUSY breaking (\eg\ gravity mediation).
The idea (already used in \Refs{AADM,CTCflavor}) is that
in the strong sector, conformal invariance 
is broken softly by SUSY breaking mass terms, giving rise
to strong non-supersymmetric dynamics at the TeV scale.
Since all scalars get massive from SUSY breaking while fermions
have chiral symmetries that forbid their masses, it is very
plausible that the strongly interacting fermions confine and
break chiral symmetries, as in QCD.
This dynamics can play a role in electroweak symmetry
breaking.
This is the conformal technicolor mechanism \cite{CTC}
in the context of SUSY, so we refer to it as
``superconformal technicolor.''%
\footnote{This name has also been used in \Refs{MinPertSCTC} for models that
do not use the conformal technicolor mechanism to break electroweak
symmetry.}

The presence of both SUSY and strong dynamics at the TeV scale
opens up many interesting phenomenological possibilities, 
and this paper only 
initiates the exploration of these ideas.
We will construct an explicit model of the strong sector
that realizes this idea, which we argue can dynamically break
electroweak symmetry.
We then investigate two different limiting regimes of the same model
that illustrate two phenomenologically distinct scenarios for
electroweak symmetry breaking.
The model has a strong conformal sector based on an $SU(2)$ gauge
group with 4 flavors, which has a strongly interacting conformal
fixed point \cite{Seiberg}.
Additional fields and interactions are required to stabilize
runaway directions in the presence of SUSY breaking.
The additional interactions and the SUSY breaking terms explicitly
break the $SU(8)$ global symmetry of the theory down to
$SU(2)_L \times SU(2)_R$, which is weakly gauged in the usual way
so that chiral symmetry breaking in the strong sector
breaks electroweak symmetry, as in technicolor.

The MSSM Higgs fields couple to the strong sector
via superpotential couplings of the form
\beq[WHiggsYukawaop]
W = \la_u H_u \scr{O}_d + \la_d H_d \scr{O}_u,
\eeq
where $\scr{O}_{u,d}$ are operators in the strong
sector with the same
electroweak quantum numbers as $H_{u,d}$.
The two different regimes of the model referred to above
correspond to different choices of $\la_{u,d}$.

In the model we construct
the operators $\scr{O}_{u,d}$ have scaling dimension
$\frac 32$, so the couplings $\la_{u,d}$ are relevant couplings
that get strong at some scale.
This scale cannot be too far from the TeV scale,
otherwise they are not important for electroweak symmetry breaking.
This amounts to a coincidence of scales, and the problem of explaining
this coincidence
is similar to the ``$\mu$ problem'' of the MSSM.
In both cases we must explain why a
relevant supersymmetric coupling
is important near the scale of SUSY breaking.
Perhaps the simplest and most elegant solution to the $\mu$ problem
is the Giudice-Masiero mechanism \cite{GM}, and we review
an extension of this mechanism \cite{CTCflavor}
that explains why the couplings
\Eq{WHiggsYukawaop} are important at the TeV scale.

\subsection{Induced Electroweak Symmetry Breaking}
We first consider the case where the couplings
\Eq{WHiggsYukawaop} are perturbative at the TeV scale.
In this case, the Higgs fields $H_{u,d}$ are ordinary perturbative
degrees of freedom below the TeV scale.
The strong sector dynamically breaks electroweak symmetry
with an order parameter $f$ that we assume is somewhat below the
value required to explain the $W$ and $Z$ masses,
\eg $f \simeq 100\GeV$.
The heavy hadrons of the strong sector are expected to have
masses of order $4\pi f \sim \mbox{TeV}$ \cite{NDA}, and the 
$SU(2)_L \times SU(2)_R$ chiral symmetry of this theory is
nonlinearly realized below this scale.
The couplings \Eq{WHiggsYukawaop} then generate a tadpole for $H_{u,d}$
in the effective potential.
This induces a VEV for $H_{u,d}$ even if $m_{H_{u,d}}^2 > 0$,
which we assume to be the case.
(In standard SUSY scenarios $m_{H_{u,d}}^2 > 0$ at high scales
and renormalization group running results in
$m_{H_u}^2 < 0$ at the TeV scale,
but more general boundary conditions at high scales can lead
to $m_{H_{u,d}}^2 > 0$ at the TeV scale.)
If we neglect the quartic terms in the potential for $H_{u,d}$,
the masses of the physical Higgs bosons are simply eigenvalues
of the quadratic terms in the effective potential, while the size
of the VEV is determined by the coefficient of the tadpole.
The Higgs mass therefore depends directly on the SUSY breaking
masses, similar to a slepton or squark mass.
The Higgs masses can easily be larger than the LEP bound with no
tuning in this scenario, giving a simple and robust
solution to the SUSY Higgs mass problem.

In this scenario
electroweak symmetry breaking is
shared by the elementary Higgs bosons and the strong sector:
\beq
v^2 = v_u^2 + v_d^2 + f^2,
\eeq
where $v = 246\GeV$.
For example, for $f \simeq 100\GeV$ we have $\sqrt{v_u^2 + v_d^2} = 225\GeV$.
Because the electroweak symmetry breaking VEV is dominantly in the
elementary Higgs fields, quark and lepton masses can arise through
conventional perturbative Yukawa couplings.
This means that there is no additional flavor problem associated
with the strong dynamics.
Of course we still have the SUSY flavor problem,
namely the squark and slepton masses and $A$ terms can be flavor-dependent.
We assume that this is addressed by one of the many possible
mechanisms in the literature.

A good precision electroweak fit can be obtained in this model.
The strong sector is based on a $SU(2)$ gauge theory with a single
technidoublet, so the corrections are not enhanced by large $N$ 
factors.
The UV contribution to the $S$ parameter is very uncertain
because this theory is very different from QCD.
The fact that the longitudinal modes of the $W$ and $Z$ are
dominantly perturbative excitations reduces the IR contribution
from the strong sector to the $S$ parameter.
The custodial symmetry breaking from $\la_u \gg \la_d$
gives positive contribution to the $T$ parameter
that also helps with the fit.
The conclusion is that we can get a good precision electroweak
fit even if we assume that the UV contribution to the $S$ parameter
is large and given by the value extrapolated from QCD.

The collider phenomenology for this model includes all of the
usual SUSY signals, together with additional signals arising from
the strong sector.
The strong sector has a relatively low scale
$4\pi f \lesssim {\rm TeV}$, which may make it more accessible than
conventional technicolor.%
\footnote{Low-scale technicolor has been previously studied,
motivated by
large $N$ technicolor theories \cite{lowscaleTC}.
However, as previously noted these theories have serious
problems with the precision electroweak fit.}
The theory below the TeV scale has 3 additional $CP$ odd states
$A_2^0$
and $H_2^\pm$ that are heavier than the other Higgs fields
and are dominantly pseudo Nambu Goldstone bosons
(PNGBs) from the strong sector.
These can be either singly produced, or pair produced from decays
of heavy resonances in the strong sector.
There are many possible signals, and we will only outline some of the
possibilities in this paper.

\subsection{Strong Electroweak Symmetry Breaking}
We then consider another possibility where there is no light Higgs
below the TeV scale.
SUSY breaking in the strong sector triggers electroweak symmetry
breaking, as in conformal technicolor.
The quark and lepton masses arise from couplings to the strong
sector of the form
\beq[dualYukawa]
\De W \sim (y_u)_{ij} Q_i u^c_j \scr{O}_u
+ (y_d)_{ij} Q_i d^c_j \scr{O}_d
+ \cdots
\eeq
This can arise in the same model we construct for the previous
scenario for a different choice of parameters.
The couplings $\la_{u,d}$ in \Eq{WHiggsYukawaop} are relevant
operators that get strong at some scale $\La_*$.
If $\La_*$ is above the SUSY breaking scale,
the elementary Higgs fields become part of the strong sector, 
and there is a dual description where the Yukawa couplings
become couplings of the form \Eq{dualYukawa}.
Below the scale $\La_*$, the operators $\scr{O}_{u,d}$ have
dimension $\frac 32$, so these operators behave like flavor-dependent
interactions in ``walking'' technicolor.%
\footnote{The use of SUSY conformal fixed points to get ``walking''
behavior of flavor couplings has been previously considered in
\Ref{technicolorfulSUSY}.}
Alternatively, the scale $\La_*$ may be naturally near the TeV scale, 
as discussed above.
In this case we do not require large Yukawa couplings at high scales.
In either case,
the couplings \Eq{dualYukawa} inherit the minimal flavor violating
structure of the Yukawa couplings, so there is no flavor
problem associated with the strong dynamics.
Of course, the SUSY flavor problem must still be addressed
by some mechanism.

The precision electroweak fit does not pose a problem for this model.
There is a contribution to the $T$ parameter from $\la_u \ne \la_d$.
If this contribution is positive (as suggested by perturbation theory)
we can get a good fit provided that the UV contribution to the
$S$ parameter from the strong sector is somewhat smaller 
(\eg\ by a factor of 2) than the QCD estimate.
We conclude that given our present state of knowledge precision
electroweak data does not strongly constrain this model.

The collider signals include the standard missing energy SUSY signals,
but not the SUSY Higgs signals.
There are technicolor-like signals associated with the strong sector.
One difference from conventional technicolor is that the strong
sector generally has no approximate parity symmetry, so the resonances
that unitarize $WW$ scattering can decay to $WWW$ as well as $WW$.

\section{The Strong Superconformal Sector
\label{sec:UVmodel}}
In this section we describe the requirements for a successful model
of the strong sector, and construct an explicit model as an existence proof.
The main issue is preventing runaway directions due to soft SUSY breaking
mass terms.

\subsection{SUSY Breaking in SUSY QCD
\label{sec:SQCD}}
The main new feature of our framework is a strongly-coupled
superconformal sector.
The simplest nontrivial 4D superconformal theory is
$SU(N_c)$ SUSY QCD with $N_f$ flavors
in the conformal window $\frac 32 N_c < N_f < 3N_c$ \cite{Seiberg}.
There is a dual description of these theories in terms of an $SU(\tilde{N}_c)$
gauge theory with $\tilde{N}_c = N_f - N_c$.
The theories with $N_f \simeq 3 N_c$ are weakly coupled,
while the models with $N_f \simeq \frac 32 N_c$ have a weakly coupled
dual description.
The models with $N_f \simeq 2N_c$ are have no weakly coupled description,
and these are 
the simplest candidates for the strong sector of our model.

Conformal symmetry is broken softly by SUSY breaking
terms in the strong sector.
We begin by reviewing what is known about
soft SUSY breaking for SUSY QCD at a conformal
fixed point \cite{CFTSUSYbreak}.
The effects of soft SUSY breaking terms are most
readily understood if we view them as 
$F$ and $D$ components of superfield couplings and flavor
gauge fields.
We write the Lagrangian in superspace as
\beq
\bal
\scr{L} &= \myint d^2 \th\, \tau \tr(W^\al W_\al) 
+ \hc
\\
&\qquad +
\myint d^4\th\, Z \left[
Q^\dagger_i e^V (e^X)^i{}_j e^Y Q^j
+ \tilde{Q}^\dagger_i e^{-V^T} (e^{\tilde{X}})^i{}_j e^{-Y}
\tilde{Q} \right].
\eal\eeq
Here $V$ and $W_\al$ are the $SU(N_c)$ gauge field and field strength,
$Q$ and $\tilde{Q}$ are the fundamental and antifundamental 
``quark'' fields;
$\tau$ is the holomorphic gauge coupling,
$Z$ is a real superfield wavefunction renormalization factor;
$X$, $\tilde{X}$, and $Y$ are background gauge superfields for
the anomaly-free $SU(N_f) \times SU(N_f) \times U(1)$ flavor symmetry.

A flavor-universal mass-squared term can be parameterized by a
$D$ term for $Z$, and a gaugino mass can be parameterized by an
$F$ term for $\tau$.
The physical gauge coupling is the lowest component of a real
superfield $R$ that is a function of $\tau$
and $Z$ \cite{Realgaugecoupling}, so these SUSY breaking terms 
perturb $R$ away from its fixed point value.
Since the fixed point is IR attractive, the SUSY breaking perturbations
scale away in the IR.
On the other hand, $D$ terms for the gauge superfields $X$, $\tilde{X}$
and $Y$ are unsuppressed in the IR because the coupling of gauge
fields in the IR is simply determined by group theory.
Scalar mass-squared terms proportional to symmetry generators
therefore scale in the IR just like in a free field theory.
Detailed elaboration of these arguments can be found in
\Ref{CFTSUSYbreak}.

This means that the only soft SUSY breaking in the
strong sector that is
naturally at the TeV scale is scalar mass-squared terms 
proportional to anomaly-free flavor generators.
There are always directions in field space where the energy
due to such mass-squared terms is negative.
The ground state will then have a large VEV along such a direction,
in which case conformal
symmetry in the strong sector is broken well above the TeV scale.%
\footnote{SUSY breaking may be communicated to the visible sector
at a scale as low as $10\TeV$.
If we assume that the soft masses at $10\TeV$ are the same
order of magnitude in the MSSM and the strong sector, and that
the anomalous dimensions that suppress the
soft terms in the strong sector are numerically small,
we may get a viable model.
We will not pursue this possibility here.}
For example a soft mass proportional to ``baryon number''
($B(Q) = -B(\tilde{Q}) = 1$) will result in a runaway
direction with either $Q \ne 0, \tilde{Q} = 0$
or $Q = 0, \tilde{Q} \ne 0$ depending on the sign of the mass-squared
term.

Generalizing from SUSY QCD, 
we see that what we would like is a strong conformal theory
with an anomaly-free flavor generator
$X$ such all of the flat directions have the same sign of the $X$ charge.
A scalar mass-squared term proportional to $X$ can then stabilize
the the vacuum at small field values.
Note that this condition is never satisfied in theories with a 
charge conjugation invariance (such as SUSY QCD).
In such theories the best we can hope for is that all
flat directions have $X = 0$, in which case a more subtle analysis
is required to determine whether the ground state is near the origin
of field space.

We can lift dangerous flat directions by introducing additional
perturbative couplings.
For example, we can lift the $B \ne 0$ flat directions in the example
above by gauging $U(1)_B$.
However, as long as the $U(1)_B$ gauge coupling is weak, this will
stabilize the VEV at a large value because the VEV goes to infinity
as the $U(1)_B$ gauge coupling goes to zero.
Such a model will have more than one scale, and will not give a
strongly-coupled model with a single scale that we are seeking.

\subsection{A Viable Model
\label{sec:viable}}
We now construct a working model in which the runaway directions
in the strong sector are lifted.
The detailed model will be described below, but we start by briefly
outlining the basic mechanism.
The strong sector us a $SU(2)$ gauge theory with $4$ flavors,
with superpotential couplings to elementary Higgs fields $H$ and 
additional singlet
fields $S$ of the from
\beq
W \sim (\la_H H + \la_S S) \Psi \Psi.
\eeq
The effect of these terms is that the flat directions of the
strong sector are replaced by flat directions of the $H$ and $S$
fields, so the problem is now to lift these flat directions.
The ``meson'' operator $\Psi\Psi$ has dimension $\frac 32$,
so the $\la$ couplings have dimension $+\frac 12$.
We will want all of the $\la$ couplings to become strong near
the TeV scale where SUSY is broken in the strong sector.
This is a coincidence problem precisely analogous to the 
``$\mu$ problem'' of the MSSM.
We will show below (in Section \ref{sec:coincidence}) that we can explain
this coincidence using a generalization of the 
Giudice-Masiero mechanism for the $\mu$ term.
Now the idea is that the couplings $\la_S$ become strong at a scale
$\La'$ somewhat above the weak scale, while the coupling $\la_H$
is still weak.
Below this scale, the theory quickly flows to a new fixed point
where $S$ is a strong operator.
In this new CFT, a universal positive soft mass for $S$ is suppressed
by a large anomalous dimension, but if the scale $\La'$ is not too
far from the TeV scale this effect can be small, and there can be
a positive soft mass at the TeV scale to stabilize the strong sector.

We now give a detailed description of the model.
It is based on a 
strong $SU(2)_{\rm SC}$
gauge theory with $4$ flavors, which has a strong
conformal fixed point as discussed above.
The anomaly-free global symmetry group is
\beq[globalsymm]
SU(2)_1 \times SU(2)_2 \times SU(2)_3 \times SU(2)_4
\times U(1)_R.
\eeq
The embedding of the electroweak gauge group in this
global symmetry will be described below.
The strongly-interacting fields  transform as
\beq[Psifields]
\bal
\Psi_1 &\sim (2, 2, 1, 1, 1)_{\frac 12},
\\
\Psi_2 &\sim (2, 1, 2, 1, 1)_{\frac 12},
\\
\Psi_3 &\sim (2, 1, 1, 2, 1)_{\frac 12},
\\
\Psi_4 &\sim (2, 1, 1, 1, 2)_{\frac 12}.
\eal\eeq
The electroweak gauge group is embedded in the global
symmetry by taking the $SU(2)_W \times U(1)_Y$
generators acting on the fields $\Psi_i$ to be
\beq[EWembed]
\!\!\!\!\!\!\!\!
T_a = \sfrac 12 
\pmatrix{\tau_a & & & \cr & 0 & & \cr & & 0 & \cr & & & 0 \cr},
\qquad
Y = \sfrac 12
\pmatrix{0 & & & \cr & -\tau_3 & & \cr & & \tau_3 & \cr & & & -\tau_3 \cr}.
\eeq
The fields $\Psi_{3,4}$ will not play a role in breaking electroweak
symmetry.
We could define \eg\ $Y = \diag(0, -\tau_3, 0, 0)$, 
but then the model has physical states with fractional
charge that we want to avoid.

The fields transform as
\beq[Psitrans]
\Psi_i \mapsto U \Psi^{\vphantom{T}}_i V_i^T,
\quad
i = 1, \ldots, 4,
\eeq
where $U \in SU(2)_{\rm SC}$, $V_i \in SU(2)_i$.
The $SU(2)_{\rm SC}$ gauge
invariant holomorphic operators are the ``meson'' fields
\beq
M_{ij} = \Psi_i^T \ep \Psi_j.
\eeq
These are $2\times 2$ matrices, transforming 
under
$SU(2)_i \times SU(2)_j$
as
\beq
M_{ij} \sim \begin{cases}
(2, 2) & for $i \ne j$,
\cr
1 & for $i = j$.
\end{cases}
\eeq

In addition to the techniquarks \Eq{Psifields}, the model contains
$SU(2)_{\rm SC}$ singlet fields $S_{ij}$ transforming under 
the global symmetries like the meson fields $M_{ij}$ above.
The theory has a superpotential 
\beq[theWUVmodel]
W = \sum_{i, j} \la_{ij} S_{ij} \Psi_i^T \ep \Psi_j. 
\eeq
The couplings $\la_{ij}$ have dimension $\frac 12$, \ie they
are relevant couplings.
We assume that there is no large
hierarchy between the $\la_{ij}$, so they all get
strong at roughly the same scale $\La_*$.

Seiberg duality tells us that 
below the scale $\La_*$ the theory flows to a new strong
fixed point.
In the ``electric'' description presented here, this fixed
point is one where the couplings $\la_{ij}$ flow to strong
fixed point values.
The dual ``magnetic'' description has gauge group
$SU(2)_{\widetilde{\rm SC}}$ and dual ``quark'' fields
$\tilde{\Psi}_i \sim (2, 2)$ under
$SU(2)_{\widetilde{\rm SC}} \times SU(2)_i$,
as well as  the ``meson'' fields
$M_{ij}$ as separate degrees of freedom.
This theory has a superpotential 
\beq[thestrongW]
\tilde{W} = \sum_{i,j} 
\left( \la_{ij} S_{ij} M_{ij}
+ M_{ij} \tilde{\Psi}_i \tilde{\Psi}_j \right).
\eeq
The first term arises from \Eq{theWUVmodel}
and the second is dynamically generated.
In this description the singlets get a mass with the
meson fields, and we can integrate them out
to get a $SU(2)_{\widetilde{\rm SC}}$
gauge theory with 8 flavors and no superpotential.
This is precisely the argument used to show that the dual of 
a Seiberg dual is the original theory, except that the couplings
$\la_{ij}$ are here allowed to violate the flavor symmetries.
This theory has a strongly-coupled IR attractive
fixed point, which shows that
the theory flows to a new fixed point below the scale where
the couplings $\la_{ij}$ become strong.

The theory below the scale $\La_*$ is pure SUSY QCD,
in which universal scalar mass-squared terms are suppressed.
However, above the scale $\La_*$ universal soft mass-squared terms
for $S$ are not suppressed, and are therefore unsuppressed at the
scale $\La_*$.
If the scale $\La_*$ is not too far above the TeV scale,
these soft mass terms can break SUSY near the TeV scale in 
the strong sector.
The effects of a universal scalar mass-squared term in the
dual description of the strong sector
below the scale $\La_*$ are discussed
in an Appendix.

In addition to scalar mass-squared terms, we can have $A$
terms for the superpotential couplings \Eq{thestrongW}.
In superspace these can be parameterized by terms
\beq
\De\scr{L} = \myint d^4\th\, (A \th^2 + \hc) S^\dagger S
\eeq
which are not suppressed by the strong dynamics above the scale
$\La_*$.
For $\La_* \sim \mbox{TeV}$ these can also be important at the
TeV scale.

Having $\La_* \sim \mbox{TeV}$
requires a coincidence of scales between the
supersymmetric relevant couplings $\la_{ij}$ and the
SUSY breaking scale.
As discussed above, this is similar to the $\mu$
problem, and we will present an explanation of it using a
generalization of the Giudice-Masiero mechanism below.

We have thus succeeded in constructing
a strong superconformal theory where all flat directions are lifted
by soft SUSY breaking.
The conformal symmetry is therefore broken by the soft SUSY breaking
in the strong sector at the scale $\MS$.
SUSY breaking gives mass to all scalars, but unbroken chiral
symmetries mean that technifermions are still massless.
It is therefore very plausible that this theory
confines and spontaneously
breaks the chiral symmetries, like QCD or technicolor.

We discuss the symmetry breaking and vacuum alignment in this model.
A useful starting point is to choose the
couplings $\la_{ij}$ and the
soft SUSY breaking terms to respect the full $SU(8)$
global symmetry of the $SU(2)_{\rm SC}$ gauge theory.
We do this by assuming universal couplings $\la_{ij}$ and a
universal positive mass-squared for the singlets in the UV.
The $U(1)_R$ symmetry is broken by $A$ terms 
of the same form as the superpotential
\Eq{thestrongW}.
In the dual description the dual techniquarks have no
superpotential interactions.
(When we include Yukawa couplings they will have
perturbative superpotential couplings with ordinary quark and
lepton superfields.)
The techniscalars all get masses,
but masses for the technifermions are forbidden by
the $SU(8)$ chiral symmetry.
A technigaugino mass is allowed because $U(1)_R$ is broken.
We expect that the strong non-supersymmetric gauge dynamics
generates a fermion condensate
\beq
\avg{\Psi^A \Psi^B} = -\avg{\Psi^B \Psi^A},
\eeq
where $A, B$ are $SU(8)$ indices.
This spontaneously breaks
$SU(8) \to Sp(8)$,
giving rise to 27 Nambu-Goldstone bosons (NGBs).

Now we turn on additional terms that
explicitly break the $SU(8)$ global symmetry down to
\beq[reducedsymm]
SU(2)_L \times SU(2)_R \times U(1)_{\tilde{Y}},
\eeq
with generators
\beq
\!\!\!\!\!\!\!\!
T_{La} = \sfrac 12 \pmatrix{ \tau_a & & & \cr & 0 & & \cr
& & 0 & \cr & & & 0 \cr},
\qquad
T_{Ra} = \sfrac 12 \pmatrix{ 0 & & & \cr & -\tau_a^T  & & \cr
& & 0 & \cr & & & 0 \cr},
\eeq
and
\beq
\tilde{Y} = \sfrac 12 \pmatrix{ 0 & & & \cr & 0 & & \cr
& & \tau_3 & \cr & & & -\tau_3 \cr}.
\eeq
This explicit breaking is accomplished by
non-universal couplings $\la_{ij}$,
and non-universal soft masses for the $S_{ij}$ and the
$\Psi_i$.
We assume that this breaking is maximal, so that there is no
larger approximate global symmetry.
This assumption is made just for simplicity, and it is also
natural in this framework to have additional approximate
global symmetries leading to pseudo Nambu-Goldstone bosons
that can have interesting phenomenological implications.

This explicit $SU(8)$ breaking determines the alignment of the
fermion condensate.
We assume that
\beq
\avg{\Psi^A \Psi^B} =
\pmatrix{0 & a 1_2 & & \cr
- a 1_2 & 0 & & \cr
& & 0 & b 1_2 \cr
& & -b 1_2 & 0 \cr},
\eeq
which breaks
\beq
SU(2)_L \times SU(2)_R &\to SU(2)_{\rm diag},
\eeq
and preserves $U(1)_{\tilde{Y}}$.
The breaks electroweak symmetry in the desired pattern,
with no pseudo Nambu-Goldstone bosons.

We now describe how this theory
generates masses for quarks and leptons.
Note that $S_{12}$ has the 
electroweak quantum numbers of 2 Higgs doublets.
We can therefore write conventional Yukawa couplings
\beq[SYukawa]
\De W = y_u (Q u^c) (S_{12})_u + y_d (Q d^c) (S_{12})_d
+ y_e (L e^c) (S_{12})_d.
\eeq
Above the scale where the couplings $\la_{ij}$ become strong,
$S_{12}$ is a conventional weakly-coupled field with dimension 1,
so the Yukawa couplings run as in the MSSM.
Below the scale where the couplings $\la_{ij}$ become strong, we
use the dual description where we integrate out $S_{ij}$ and
the meson fields $M_{ij}$, and we obtain the superpotential
\beq[ETC]
\!\!\!\!\!\!\!\!\!\!
\De W = \frac{1}{\la_{12}} \left[
(y_u)_{ij} Q_i u^c_j (\tilde{\Psi}_1 \tilde{\Psi}_2)_u
+ (y_d)_{ij} Q_i d^c_j (\tilde{\Psi}_1 \tilde{\Psi}_2)_d
+ (y_e)_{ij} L_i e^c_j (\tilde{\Psi}_1 \tilde{\Psi}_2)_d
\right].
\eeq
Note that these interactions have minimal flavor violating structure
inherited from the Yukawa couplings \Eq{SYukawa}.
The operators $\tilde\Psi \tilde\Psi$ have dimension $\frac 32$
in the new fixed point, so we have \eg
\beq
m_t \sim y_t(\La_*) v \left( \frac{\mbox{TeV}}{\La_*} \right)^{1/2}.
\eeq
We see that
the quark masses have a mild suppression even if $\La_* > \mbox{TeV}$.

\subsection{A Model with a Light Higgs
\label{sec:lightHiggs}}
As we have described it, this models has no light
Higgs field below the SUSY breaking scale.
Since $S_{12}$ contains the MSSM Higgs fields, it is 
easy to modify the theory to have a light Higgs:\
we simply choose the
coupling $\la_{12}$ to be smaller than the others.
We assume that the other couplings $\la_{ij}$ have the same
order of magnitude,
and get strong at a single scale $\La_* \gsim \mbox{TeV}$.

In the ``electric'' description of the theory, the strong
Yukawa couplings $\la_{ij}$ approach a strong fixed point,
while $\la_{12}$ remains weak.
In the dual ``magnetic'' description 
the strong $\la_{ij}$ turn into mass terms
of order $\La_*$, while $\la_{12}$ is a smaller mass term.
After integrating out the masses of order $\La_*$,
the dual superpotential is
\beq
\tilde{W} = \la_{12} S_{12} M_{12} + M_{12} \tilde{\Psi}_1 \tilde{\Psi}_2.
\eeq
In this description there is an additional light
$SU(2)_{\rm SC}$ singlet field $M_{12}$,
but it has a strong superpotential coupling to the dual techniquarks,
and should be viewed as part of the strong sector.
In either description, assuming that $\la_{12}$ is small at the SUSY
breaking scale, it will give rise to a weak coupling of the elementary
Higgs fields in $S_{12}$ to the strong dynamics.
This strong dynamics can still have 
the symmetry structure described above,
and it is equally plausible that it is spontaneously broken in the
same pattern.
This is all we need for the low-energy dynamics we are trying to 
achieve.

\subsection{Coincidence Problem
\label{sec:coincidence}}
We now discuss the coincidence between the SUSY breaking
scale and the scale where the couplings $\la_{ij}$ become
strong.
We describe how this can happen in an extension of the Giudice-Masiero
mechanism \cite{CTCflavor}.
We assume that SUSY is broken in
a hidden sector at high scales, and is communicated
to the visible sector by higher-dimension operators.
The hidden sector contains a
gauge singlet superfield $X$ with $\avg{F_X} \ne 0$,
and higher dimension interactions that connect the hidden
and the visible sector are suppressed by a
scale $M$.
We then write all possible
higher-dimension operators coupling
$X$ to the visible sector fields, \eg
\beq
\bal
\De\scr{L}_{\rm eff} 
&\sim \myint d^2 \th\, \frac{1}{M} X W^\al W_\al + \hc
\\
&\qquad
+ \myint d^4\th \left[
\frac{1}{M} (X + X^\dagger) Q^\dagger Q
+ \frac{1}{M^2} X^\dagger X Q^\dagger Q
\right]
\\
&\qquad
+ \myint d^4\th\, \left[ \frac{1}{M} X^\dagger H_u H_d 
+ \frac{1}{M^2} X^\dagger X H_u H_d + \hc \right].
\eal
\eeq
These terms generate respectively gaugino masses, $A$ terms, 
scalar mass terms, the $\mu$ term, and the $B\mu$ term,
all of order
\beq
\MS \sim \frac{\avg{F_X}}{M}.
\eeq
Note also that the soft terms in the MSSM and the strong sector are
generated at the same scale in this mechanism.
One well-motivated choice is to take $M$ of order the Planck scale,
in which case one must also take into account supergravity corrections,
but they do not change this result \cite{GM}.
The main shortcoming of this mechanism is that it does not address
the SUSY flavor problem, which is why the soft masses are flavor
diagonal.
On the other hand, models that address the SUSY flavor problem
require significant complications to solve the $\mu$ problem,
and it is not  obvious which is preferred.

In the model above, the couplings $\la_{ij}$ have mass dimension
$\frac 12$, so the
problem is to naturally generate $\la_{ij} \sim \MS^{1/2}$.
This occurs naturally if the hidden sector contains a field
$Y$ with
\beq[Ycond]
\avg{Y} \sim \avg{F_X}^{1/2},
\qquad
\avg{F_Y} \lsim \avg{F_X}^{1/2} \MS.
\eeq
The couplings $\la_{ij}$ can then be generated by
\beq
\De W = \frac{c_{ij}}{M^{1/2}} Y S_{ij} \Psi_i \Psi_j.
\eeq
The second condition in \Eq{Ycond} is required to ensure that 
this does not generate large $A$ terms.
For example, \Ref{CTCflavor} shows that a hidden sector with superpotential
\beq
W = \ka X + \frac{1}{M} Y^4
\eeq
has the desired features, even if supergravity effects are included.
In this model $\ka$ sets the scale of the VEVs.
The fact that $Y$ and not $X$ couples to the operator $S_{ij} \Psi_i \Psi_j$
can be enforced by symmetries, \eg\ discrete $R$ symmetries.
This requires only a modest generalization of the hidden
sector, and we believe it is natural in the aesthetic as well
as the technical sense.

\subsection{Discrete Symmetries
\label{sec:discrete}}
We now discuss the discrete symmetries of the strong sector described
above.
Because the theory is based on a $SU(2)$ gauge group,
there is no spacetime parity symmetry.
$CP$ is still a good symmetry (assuming that the soft SUSY breaking
parameters are real).
As discussed above, the theory has a $SU(8)$ flavor group that
is explicitly broken down to
$SU(2)_L \times SU(2)_R \times U(1)_{\tilde{Y}}$.
The $SU(8)$ symmetry includes transformations that interchange the 
techniquarks charged under $SU(2)_L$ and $SU(2)_R$, but these
are broken by (for example) different soft masses for the
$L$ and $R$ techniscalars.
The scale of confinement and chiral symmetry breaking is
given by these same SUSY breaking masses (assuming there is
no hierarchy among them),
so in general there
is no approximate symmetry that interchanges $SU(2)_L$ 
and $SU(2)_R$.

This means that the hadronic states of the strong sector
are classified by their quantum numbers under the
custodial $SU(2)$ (``isospin'') and $CP$ only.
This has phenomenological implications for the heavy resonances
at the TeV scale.
The 3 Nambu-Goldstone bosons $\pi$ that arise from the symmetry breaking
pattern $SU(2)_L \times SU(2)_R \to SU(2)$ have scattering
amplitudes that grow with energy, and on general grounds we
expect this to be unitarized by strong resonances at the TeV scale.
Because there is no parity symmetry, these resonances can decay
to $\pi\pi\pi$ as well as $\pi\pi$.
When we couple this theory to the standard model, the longitudinal
$W$ will have an admixture of the $\pi$ fields, and so the 
strong resonances can decay to $WWW$ as well as $WW$.
This can provide an interesting signal of this class of models
that distinguish it from conventional technicolor models.

The absence of a parity symmetry is very general in the class
of theories we are considering.
In any gauge theory, scalars belonging to different irreducible
multiplets will in general have different masses, and there will
be no discrete symmetry interchanging them.
In a non-SUSY technicolor theory, the only relevant terms that
can break symmetries of this kind are mass terms.
Mass terms for $SU(2)_L$ and $SU(2)_R$ fermions are allowed
only in theories based on the $Sp(2N_c)$ strong gauge groups
(including $SU(2)$).
A non-SUSY example without parity is therefore minimal conformal
technicolor based on an $SU(2)$ strong gauge group with fermion
mass terms at the TeV scale \cite{minCTC}.

\section{Induced Electroweak Symmetry Breaking
\label{sec:TCinduced}}
We now consider the effective theory below the scale of confinement
and chiral symmetry breaking in the strong sector.
This theory controls the most prominent features of the phenomenology
of these models, and depends only on a few qualitative features
of the strong sector.
We start with the case where the elementary Higgs fields are
weakly coupled to the strong sector and are therefore present as
light fields in the effective theory.

\subsection{Low Energy Effective Theory of the Strong Sector}
We first enumerate the assumptions about the strong sector that
define the low-energy theory that describes the phenomenology.
We assume that the strong sector has a
$SU(2)_L \times SU(2)_R$
global symmetry that is spontaneously broken down to $SU(2)_V$ with
order parameter $f$.
The $SU(2)_L \times SU(2)_R$ symmetry is then weakly gauged by
$SU(2)_W \times U(1)_Y$ in the standard way (see previous section),
so that the electroweak gauge group is broken down to $U(1)_{\rm EM}$
with an approximate custodial symmmetry.
The low-energy theory of the strong sector
then has 3 Nambu-Goldstone bosons with
decay constant $f$.
The effective theory breaks down
at the scale $\La \sim 4\pi f$, which we identify with the scale
of confinement and chiral symmetry breaking in the strong sector
\cite{NDA}.
We assume that $f$ is somewhat smaller than what is required to
explain the $W$ and $Z$ masses, \eg\ $f \simeq 100\GeV$.
In this case, the scale $\La \sim \mbox{TeV}$ is still larger
than the $W$ and $Z$ masses, so it makes sense to describe
electroweak symmetry breaking within the effective theory below
the scale $\La$.

We assume that the strong sector is coupled to the Higgs fields
of the MSSM 
by Yukawa couplings of the form
\beq[strongHcoup]
\De\scr{L} = \la_u H_u \Om_u^\dagger + \la_d H_d \Om_d^\dagger,
\eeq
where $\Om_{u,d}$ are scalar operators with the same electroweak
quantum numbers as $H_{u,d}$.
To keep track of the custodial symmetry in the strong sector, we define
the $2 \times 2$ matrices
\beq
\Om = \pmatrix{\Om_d & \Om_u \cr},
\eeq
transforming as
\beq
\Om \mapsto L \Om R^\dagger.
\eeq
We assume that $\Om$ is an order parameter for electroweak
symmetry breaking, \ie
\beq
\avg{\Om} \propto 1_2.
\eeq
Similarly, we define
\beq[eq:TSpurions]
\scr{H} = \pmatrix{ H_d & H_u \cr},
\qquad
\la = \pmatrix{\la_d & 0 \cr 0 & \la_u \cr},
\eeq
transforming as
\beq[eq:SpurionTransformation]
\scr{H} \mapsto L \scr{H}\tilde{R}^\dagger,
\qquad
\la \mapsto \tilde{R} \la R^\dagger.
\eeq
where $\tilde{R}$ is a $SU(2)_{\tilde{R}}$ transformation.
Gauged $U(1)_Y$ transformations correspond to 
\beq
R = \tilde{R} = e^{-i\th \tau_3/2}.
\eeq
In particular, the spurion $\la$ is gauge invariant.
This implies that
\beq
\scr{H} \la \mapsto L (\scr{H} \la) R^\dagger,
\qquad
\la^\dagger \la \mapsto R (\la^\dagger \la) R^\dagger
\eeq
are spurions that can break custodial symmetry of the strong
sector in the effective theory.

The $SU(2)_L \times SU(2)_R$ symmetry is nonlinearly realized
by fields $\Si(x) \in SU(2)$ 
transforming as
\beq
\Si = e^{2i\Pi / f} \mapsto L \Si R^\dagger.
\eeq
The kinetic term and leading interaction term for these fields
are contained in the effective coupling
\beq
\De \scr{L}_{\rm eff} = \frac{f^2}{4} \tr(D^\mu \Si^\dagger D_\mu \Si)
+ \hc
\eeq

To define the terms arising from the couplings \Eq{strongHcoup}
to the elementary Higgs fields we define the normalization of the
couplings $\la_{u,d}$.
As discussed in the previous section, these are relevant interactions
above the scale $\La$, and are therefore naturally viewed as dimensionful.
In order to discuss their effects in the low-energy theory, we find it
most convenient to make them dimensionless by multiplying by appropriate
powers of $\La$.
This is a measure of the dimensionless strength of these couplings
at the scale $\La$ where we match onto the low-energy effective theory.
We then scale these couplings so that that $\la_{u,d} \sim 4\pi$
corresponds to strong coupling at the scale $\La$.
This is the normalization appropriate to dimensionless Yukawa couplings.

We now consider the terms with no derivatives,
\ie\ the potential terms.
Expanding in powers of the elementary Higgs fields, we have
\beq[TCHiggspot]
V_{\rm eff} = \frac{\La^4}{16\pi^2} \biggl[
\frac{c_1}{\La} \tr(\scr{H} \la \Si^\dagger)
+ \hc
+ \scr{O}\left( (\scr{H} \la / \La)^2 \right) \biggr].
\eeq
The size of these terms can be understood 
from the fact that they become strong at the scale $\La$
in the limit $\scr{H} \to f$, $\la \to 4\pi$.
This implies that the dimensionless couplings in \Eq{TCHiggspot}
are order 1.

We focus on the predictive scenario
where $\scr{H} \la/ \La \ll 1$.
The expansion is then in powers of
\beq[epsilondefn]
\ep = \frac{v \la}{\La} = 
\frac{1}{\La} \pmatrix{\la_u v_u & 0 \cr 0 & \la_d v_d \cr}.
\eeq
In order to stabilize the Higgs VEV at this value, we need the
soft masses for the Higgs fields to satisfy
\beq[mHinequality]
m_H^2 \gg \frac{\la^2}{16\pi^2} \La^2.
\eeq
We assume that $m_{H_u}^2, m_{H_d}^2 > 0$
so that the VEVs for the Higgs fields are induced by the
linear term in \Eq{TCHiggspot}.
Neglecting quartic terms and the $B\mu$ terms in the Higgs potential, 
minimizing the potential gives
\beq[SimpleVEV]
m_H^2 \sim \frac{\la}{4\pi}\, \frac{f}{v} \, \La^2
\sim \ep \, \frac{f^2}{v^2} \, \La.
\eeq
This is consistent with \Eq{mHinequality} provided $\ep \ll 1$.
The parameter space of this scenario will be explored in detail
below,
including the boundary of
the region where the expansion is under theoretical control.
An example of a viable choice of parameters to keep in mind is 
\beq
f = 100\GeV,
\qquad
\tan\be = 10,
\qquad
m_h = 120\GeV,
\eeq
which corresponds to $v_u = 224\GeV$, $v_d = 22\GeV$, and
$\la_u/4\pi \sim 0.03$.

\subsection{The Scalar Sector}
We now consider the scalar sector of the effective theory,
including all mixing effects.
This sector depends on 6 couplings:
the soft masses
$m_{H_u}^2, m_{H_d}^2, B\mu$,
the scale $f$,
and  the effective couplings in \Eq{TCHiggspot}
\beq
\ka_{u,d} = \frac{c_1 \La^3}{16\pi^2} \la_{u,d}.
\eeq
We can redefine the fields to make
$\ka_{u,d} > 0$.
The sign of $B\mu$ is then physically meaningful.
Because the VEV $v$ is measured, the scalar sector has 5 parameters,
which we can take to be \eg
\beq
\tan\be,\ f,\ m_{H_u}^2,\ m_{H_d}^2,\ B\mu.
\eeq

We parameterize the scalar fields as
\beq
H_u = \pmatrix{H_u^+ \cr \displaystyle \frac{1}{\sqrt{2}}
(v_u + h_u^0 - i A_u^0)},
\qquad
H_d = \pmatrix{\displaystyle \frac{1}{\sqrt{2}}
(v_d + h_d^0 + i A_d^0) \cr H_d^- \cr},
\eeq
and
\beq
\Pi = \frac{1}{\sqrt{2}} \pmatrix{
\pi^0 / \sqrt{2} & \pi^+ \cr
\pi^- & -\pi^0 / \sqrt{2} \cr}.
\eeq
We define fields perpendicular to the eaten Goldstones by
\beq
\pmatrix{ A_d^0 \cr A_u^0 \cr \pi^0 \cr} = U
\pmatrix{A_h^0 \cr A_\pi^0 \cr G^0 \cr},
\qquad
\pmatrix{ H_d^\pm \cr H_u^\pm \cr \pi^\pm \cr}
= U \pmatrix{H_h^\pm \cr H_\pi^\pm \cr G^\pm \cr},
\eeq
where
\beq
U = \pmatrix{s_\be & -c_\ga c_\be & -s_\ga c_\be \cr
c_\be & c_\ga s_\be  & s_\ga s_\be \cr
0 & s_\ga  & -c_\ga \cr},
\eeq
with
\beq
\tan\be = \frac{v_u}{v_d},
\qquad
\tan\ga = \frac{v_h}{f},
\qquad
v_h = \sqrt{v_u^2 + v_d^2}.
\eeq
The Goldstone modes $G^0, G^\pm$ are massless eigenstates
orthogonal to the other modes, so we have $2 \times 2$
mass matrices for the $CP$ even, $CP$ odd neutral,
and $CP$ odd charged scalars.
For the $CP$ even scalars, the mass matrix is
\beq
\!\!\!\!\!\!\!\!
M^2_{h_u^0, h_u^0}
&= m_{H_u}^2 - 2 m_Z^2 (s_\be^2 - \sfrac 14) s_\ga^2,
\\
M^2_{h_u^0, h_d^0}
&=  -B\mu - m_Z^2 s_\be c_\be s_\ga^2,
\\
M^2_{h_d^0, h_d^0}
&= m_{H_d}^2 - 2 m_Z^2 (4 s_\be^2 - 3) s_\ga^2.
\eeq
For the $CP$ odd neutral scalars, we have
\beq
\!\!\!\!\!\!\!\!
M^2_{A_h, A_h}
&= m_{H_u}^2 c_\be^2 + m_{H_d}^2 s_\be^2
+ 2 B \mu s_\be c_\be
- \sfrac 12 m_Z^2 (s_\be^2 - c_\be^2)^2 s_\ga^2,
\\
M^2_{A_h, A_\pi}
&= \frac{1}{c_\ga}
\left[ (m_{H_u}^2 - m_{H_d}^2) s_\be c_\be 
+ B\mu (s_\be^2 - c_\be^2) - m_Z^2 (s_\be^2 - c_\be^2) s_\ga^2) \right],
\\
M^2_{A_\pi, A_\pi}
&= \frac{1}{c_\ga^2}
\left[ m_{H_u}^2 s_\be^2 + m_{H_d}^2 c_\be^2
- 2 B\mu s_\be c_\be
+ \sfrac 12 m_Z^2 (s_\be^2 - c_\be^2) s_\ga^2 \right].
\eeq
For the charged scalars we have
\beq
\!\!\!\!\!\!\!\!\!\!\!\!\!
M^2_{H_h^\pm, H_\pi^\mp}
&= M^2_{A_h, A_h},
\\
M^2_{H_h^\pm, H_\pi^\mp}
&= \frac{1}{c_\ga} \left[
(m_{H_u}^2 - m_{H_d}^2) s_\be c_\be 
+ B\mu (s_\be^2 - c_\be^2) + m_Z^2 s_\be c_\be
(s_\be^2 - c_\be^2) s_\ga^2) \right],
\\
M^2_{H_\pi^\pm, H_\pi^\mp}
&= M^2_{A_\pi, A_\pi}.
\eeq
We define the mass eigenstates by
\beq
\pmatrix{h^0 \cr H^0 \cr}
&= \pmatrix{\cos\al & \sin\al \cr
-\sin\al & \cos\al \cr}
\pmatrix{h_u \cr h_d \cr},
\\
\pmatrix{A_1^0 \cr A_2^0 \cr}
&= \pmatrix{\cos\al_A & \sin\al_A \cr
-\sin\al_A & \cos\al_A \cr}
\pmatrix{A^0_h \cr A^0_\pi \cr},
\\
\pmatrix{H_1^\pm \cr H_2^\pm \cr}
&= \pmatrix{\cos\al_H & \sin\al_H \cr
-\sin\al_H & \cos\al_H \cr}
\pmatrix{H_h^\pm \cr H_\pi^\pm \cr},
\eeq
where
\beq
\tan 2\alpha
= \frac{2M^2_{h_u^0 h_d^0}}{M^2_{h_u^0 h_u^0}-M^2_{h_d^0 h_d^0}},
\eeq
{\it etc\/}.

We can understand the qualitative features of the scalar spectrum
by considering a simplified
limit where $B\mu = 0$ and we neglect the quartic interactions
which give rise to the terms proportional to $m_Z^2$ in the mass
matrices.
In this limit, $h_{u,d}^0$ are mass eigenstates with mass
$m_{H_{u,d}}$, and the masses of the $CP$-odd scalars are
(for $f \ll v$)
\beq
m_{A^0_1}^2 &= m_{H^\pm_1} = \frac{m_{H_u}^2 m_{H_d}^2}
{m_{H_u}^2 s^2_\be + m_{H_d}^2 c^2_\be},
\\
m_{A^0_2}^2 &= m_{H^\pm_2} = \frac{1}{c_\ga^2}
(m_{H_u}^2 s^2_\be + m_{H_d}^2 c^2_\be),
\eeq
with mixing angle
\beq
\al_{A,H} =
-\frac{m_{H_u}^2 - m_{H_d}^2}
{m_{H_u}^2 s_\be^2 + m_{H_d}^2 c_\be^2} s_\be c_\be c_\ga
\sim \frac{f}{v}.
\eeq
We see that for $c_\ga = f/v \ll 1$ the $CP$ odd mass eigenstates
$A_1^0$ and $H_1^\pm$ are dominantly elementary Higgs particles.
The states $A_2^0$, $H_2^\pm$ have masses $\sim v/f$ times larger
and are dominantly PNGBs from the strong sector.
The mixing between these two sets of states is of order $f/v$.
Using $c_\ga \sim f/v$ and the equation for $v$ (see 
\Eq{SimpleVEV}), we see that the condition that the heavy
fields have masses below the scale $\La$ is equivalent
to the condition $\ep \ll 1$.

\begin{figure}[h]
\begin{center}
\includegraphics[width=7.5cm]{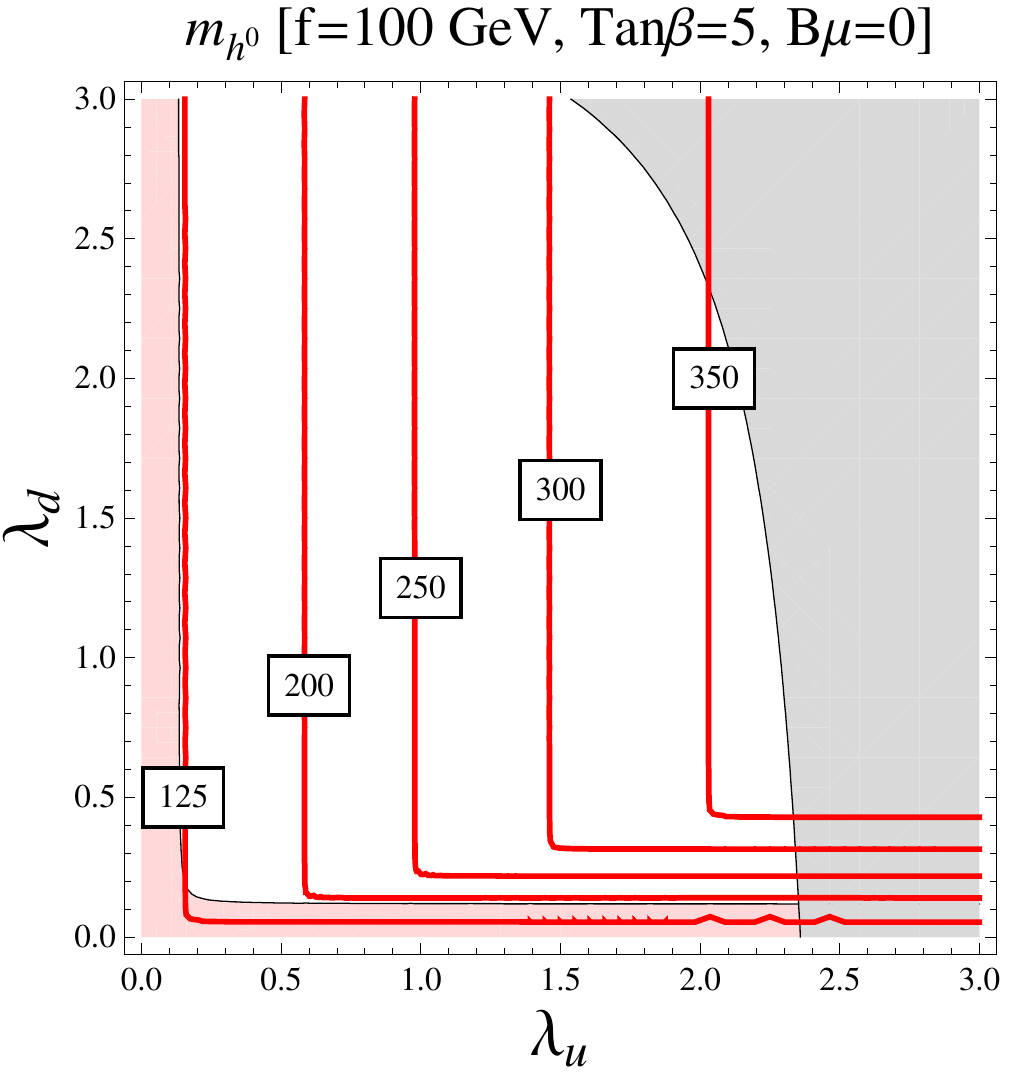}
\includegraphics[width=7.5cm]{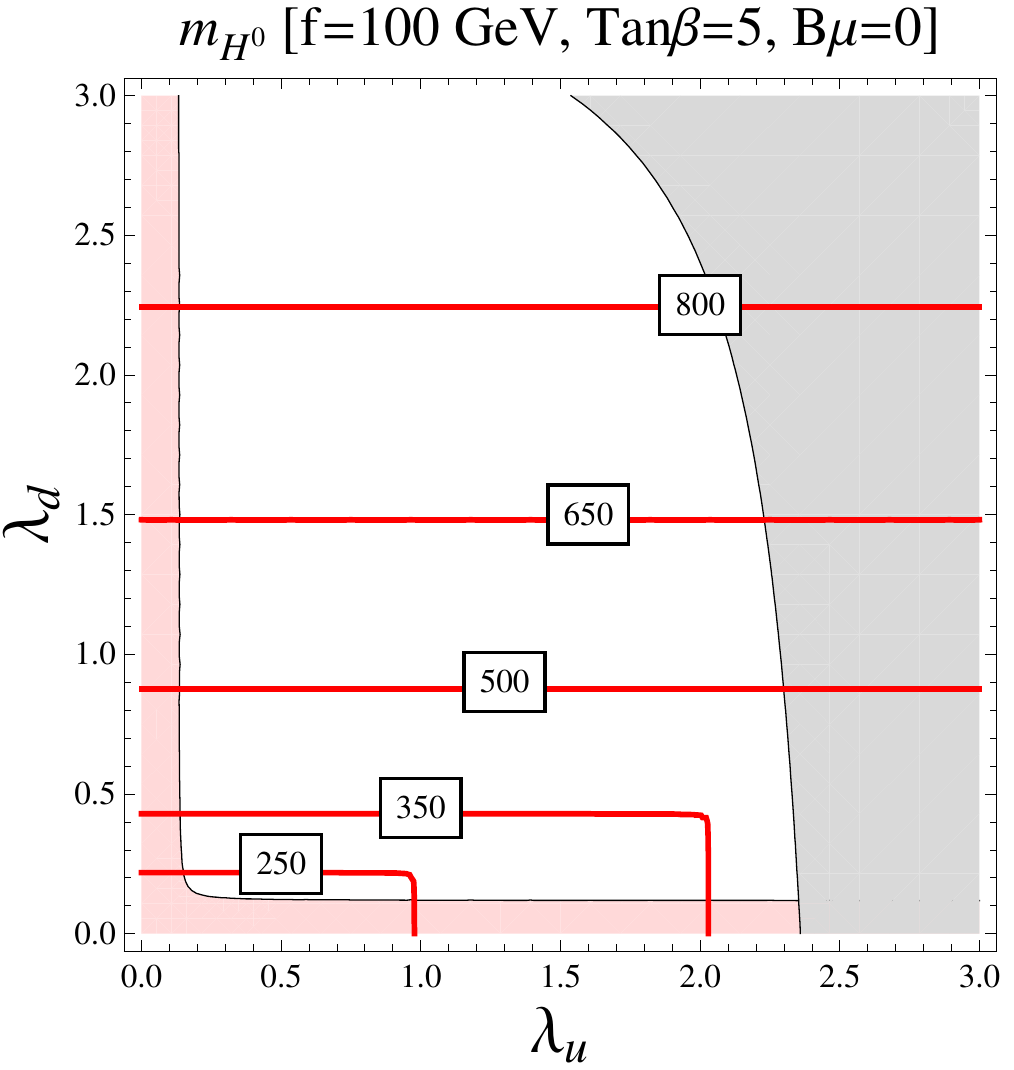}
\caption{Left panel: Masses (in GeV) for the  light $CP$ even Higgs $h^0$.
Right panel: Masses for the heavy $CP$ even Higgs $H^0$.  
The model has $f = 100\GeV$, $\tan\be = 5$, and $B\mu = 0$,
so all masses are a function of $\la_{u,d}$ normalized so that
$c_1 = 1$ in \Eq{TCHiggspot}.
The upper grey shaded region is where the perturbative expansion
breaks down, and the lower pink region is where the charged Higgs
contribution to $b \to s \ga$ comes into tension with experiment.}
\label{fig:Masses1}
\end{center}
\end{figure}

\begin{figure}[h]
\begin{center}
\includegraphics[width=7.5cm]{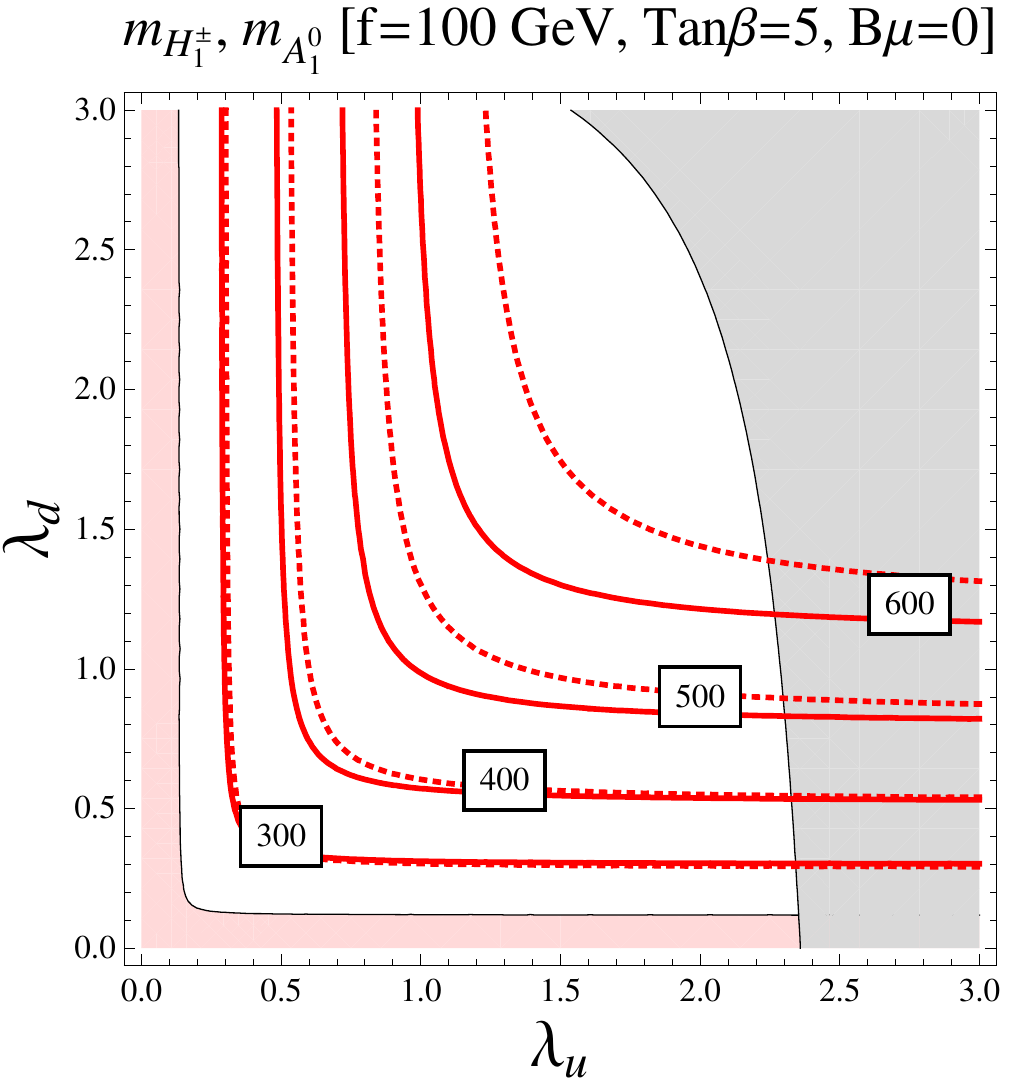}
\includegraphics[width=7.5cm]{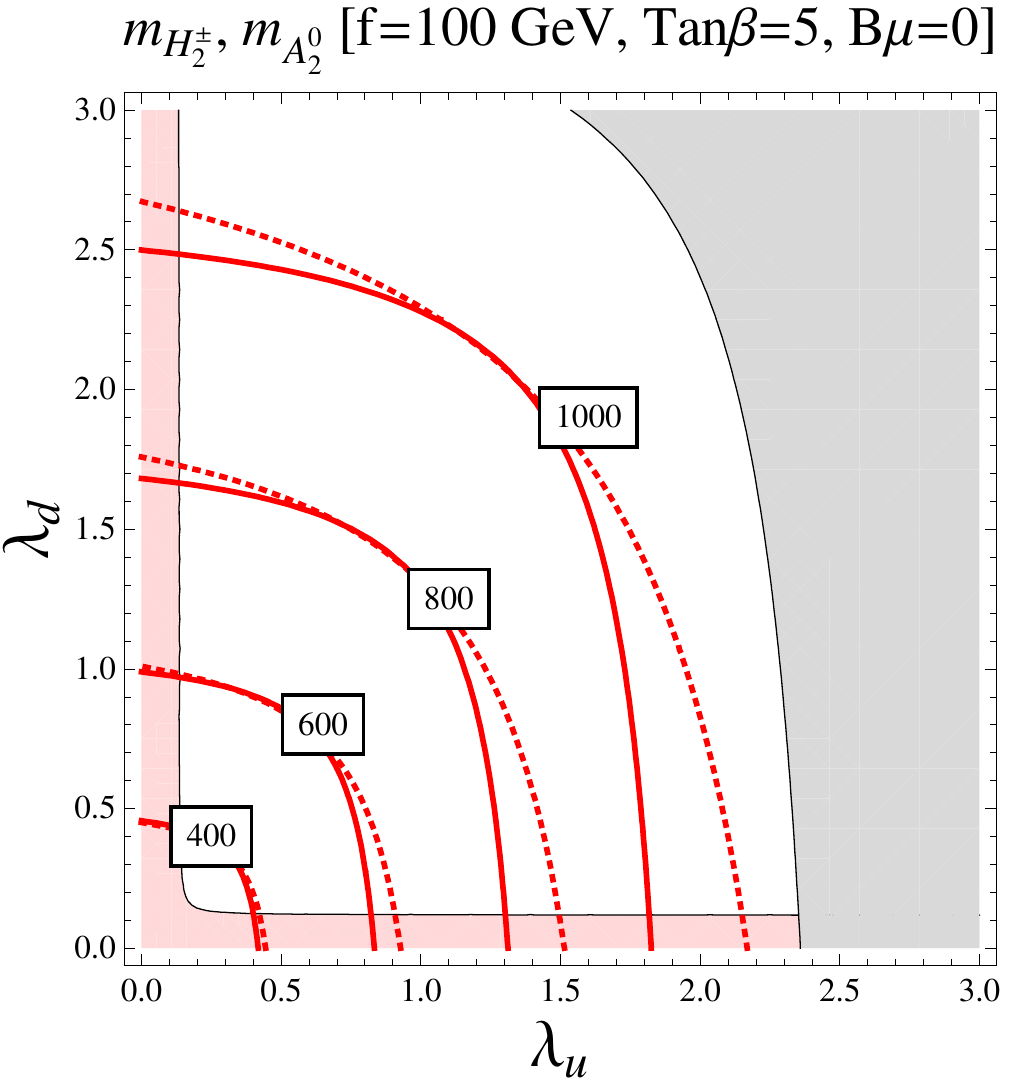}
\caption{Left panel: Masses (in GeV) for light $CP$ odd Higgs particles.
Solid lines denote $A_1^0$, dotted lines denote $H_1^\pm$.
Right panel: Likewise for $A_2^0$ and $H_2^\pm$.
The shaded regions are as in Fig.~1.}
\label{fig:Masses2}
\end{center}
\end{figure}

Some spectra including the full potential effects are illustrated
in Figs.~\ref{fig:Masses1}, \ref{fig:Masses2}.
The low-energy expansion breaks down when the heavy scalars have
mass of order $\La$, indicated by the upper grey shaded region.
For light charged Higgs scalars, there is a constraint from
$b \to s \ga$ that is indicated by the lower pink shaded region
(see \eg\ \cite{ChargedHiggs2HDM}).
Here we have neglected possible destructive interference from
Higgsino contributions that may weaken the bound.
We see that this constraint prefers somewhat heavier $h^0$
masses, but does not rule out much of the parameter space.

The couplings of these fields to standard model states are
straightforward to work out using the formulas above.
The qualitative features are that the new heavy states
$A_2^0$ and $H_2^\pm$ mix with the light Higgs fields
at order $f/v$.
These fields will therefore couple most strongly to the
heaviest standard model particles, but with a strength
suppressed by $\scr{O}(f/v)$ compared to the lighter
MSSM Higgs fields with the same quantum numbers.

\subsection{Precision Electroweak Fit
\label{sec:PEWinduced}}
We now discuss the
precision fit for the case of induced EWSB.
The only couplings of the strong sector to the 
MSSM are via electroweak gauge couplings and Higgs couplings.
The most important electroweak corrections are therefore
the oblique corrections parameterized by the electroweak parameters
$S$ and $T$, and the corrections to the $Z\bar{b}b$ vertex.

We begin with the $S$ parameter.
The physics above the confinement scale $\La$ in the strong
sector gives rise to a UV contribution to the $S$ parameter
that can be parameterized by the effective Lagrangian coupling
\beq
\Delta \lag_{\rm eff} = \frac{g g'}{16 \pi} S_{\rm UV}
\tr(\Sigma^\dagger W^3_{\mu \nu} \Sigma B^{\mu \nu}).
\eeq
The first point to make is that the strong sector need not
have either a large number of technicolors $N_{\rm TC}$
or technidoublets $N_{\rm TD}$,
which would enhance the $S$ parameter.
Traditional technicolor models generally require
both $N_{\rm TC}$ and $N_{\rm TD}$ to be large to be embedded
into extended technicolor.
Since the quark and lepton masses arise from elementary Higgs
fields, there is no reason for these parameters to be large.
For example, the theory in Section~\ref{sec:UVmodel} has
$N_{\rm TC} = 2$ and $N_{\rm TD} = 1$.

The size of the UV contribution to the $S$ parameter
is very uncertain.
Na\"\i{}ve dimensional analysis (NDA) \cite{NDA} tells us that
\beq[SNDA]
S_{\rm UV} \sim \frac{1}{\pi}.
\eeq
This is the same estimate as in technicolor theories, even though $f < v$
in this theory.
There is no suppression by powers of $f/v$ because we are in the
regime where $\La \sim 4\pi f \gg m_W$.
In the effective theory below the scale $\La$, $S$ is a dimensionless
quantity that is independent of the scale $f$.
In terms of a resonance saturation picture, $S_{\rm UV} \sim f^2 / m_\rho^2$
where $m_{\rho}$ is the resonance mass;
since $m_\rho \sim f$, the result is independent of $f$.

The $S$ parameter in traditional technicolor theories can be estimated
by scaling from QCD.
Using large-$N_c$ scaling,
one obtains \cite{SQCD}
\beq[SQCDeq]
S_{\rm UV}({\rm QCD}) \simeq 0.25 \,
\frac{N_{\rm TC}}{3} N_{\rm TD} .
\eeq
Note that this is consistent with the NDA estimate \Eq{SNDA}.
But \Eq{SQCDeq} is better than an order of magnitude estimate
only if the spectrum and couplings
at the strong scale $\La$ are similar to QCD.
However, the present theory is supersymmetric and conformal above
the scale $\La$, and there is no reason to believe that this is the case.
In fact, it has been argued that theories that are conformal
above the scale $\La$ have a significantly reduced $S$
parameter \cite{sundrumhsu}.
There is also some support for a smaller $S$ parameter from
lattice simulations.
A recent lattice simulation with $N_c = 3$, $N_f = 6$
found that the $S$ parameter \emph{per electroweak doublet}
is reduced compared to QCD by a factor between $0.3$ and $0.6$
\cite{Slattice}.
This theory is not conformal, but this at least
emphasizes the large uncertainty in the $S$ parameter
from strongly coupled electroweak symmetry breaking sectors.

Our theoretical understanding of the $S$ parameter in
strongly coupled theories is very poor.
For example, there is no rigorous theoretical understanding 
of even the sign of the $S$ parameter in QCD, where many rigorous
inequalities are known \cite{QCDinequalities}.
Data tells us that $S > 0$ in QCD, and Weinberg sum rules
relate this to basic features of the hadron spectrum.
In QCD, the $S$ parameter can be well approximated by the
contributions from the $\rho$ and $a_1$ vector resonances,
and the positivity of $S$ follows from the fact that
$m_{a_1} > m_\rho$.
However, the present theory has no parity symmetry and there
is no symmetry distinction between the analogs of the $\rho$
and $a_1$.
If vector meson dominance holds in the present theory, the sign
of $S$ will depend on whether the couplings of the lightest
resonance are more like the $\rho$ or the $a_1$.
The breaking of parity symmetry depends on the SUSY breaking masses,
so the UV contribution to $S$ will change by $\scr{O}(100\%)$
as these parameters are varied.
It is very plausible that
there are choices of parameters where it is significantly reduced,
perhaps even negative.
On the other hand,
5D AdS models can be interpreted as ``holographic'' descriptions
of large-$N$ conformal field theories, and in these theories
$S$ is positive whenever it is calculable \cite{Sholographic}.
In perturbation theory, $S$ is generally positive unless special 
representations and couplings are chosen \cite{SnegativePT}.
Perhaps these are hints that nature prefers $S > 0$.

In this paper, we will us the QCD value for the UV contribution
to the $S$ parameter as a benchmark, allowing us to make plots
and gauge the impact of precision electroweak data on this model.
As argued above, this is a conservative benchmark.
We will see that we can get a good precision electroweak fit even
with these assumptions, which means that precision electroweak
data is not a strong constraint on this class of models.

There is an additional contribution to the $S$ parameter coming from
states below the scale $\La$, the 
Nambu-Goldstone bosons in the strong sector 
and the elementary Higgs fields.
These mix at order $f/v$, but given the large uncertainties
in UV contribution, we will give the result neglecting these effects.
For large $\tan\be$, electroweak symmetry breaking is dominated 
by $H_u$, while $H_d$ is decoupled, and we obtain
\beq
S_{\rm IR} \simeq \frac{1}{12\pi} 
\left[ \ln \frac{m_h^2}{m_{h,{\rm ref}}^2}
+ \ln \frac{\La^2}{m_{\pi}^2} \right].
\eeq
The first term is the standard model Higgs contribution, 
while the second is the contribution from the composite
pseudo Nambu-Goldstone bosons in the strong sector.
The first contribution is suppressed for light Higgs masses
as usual, while the second is suppressed compared to conventional
technicolor theories because the $\pi$ fields are heavy.
This means that the IR contribution to the $S$ parameter is
significantly reduced compared to ordinary technicolor.

We now turn to the $T$ parameter.
The couplings $\la_{u,d}$ in \Eq{strongHcoup} violate custodial $SU(2)$
for $\la_u v_u \ne \la_d v_d$, so the $T$ parameter depends on
adjustable parameters parameters.
This can help give a good precision electroweak fit,
as we will see.

In order to contribute to the $T$ parameter, we need a spurion
transforming as an isospin 2 representation of custodial $SU(2)$.
The spurions $\la^\dagger \la$ and $\scr{H}\la$ are both
isospin 1 (see \Eq{eq:SpurionTransformation}), so the leading contribution
to the $T$ parameter is quadratic in these spurions.
The spurion $\la^\dagger \la$ always comes from diagrams with
a loop of elementary Higgs fields, so we have
\beq
\scr{L}_{\rm eff} \sim \frac{\La^4}{16\pi^2}
\, \scr{F} \left( \frac{D_\mu}{\La}, 
\frac{\la^\dagger \la}{16\pi^2},
\frac{\scr{H}\la}{\La} \right),
\eeq
where $\scr{F}$ is an order-1 function of dimensionless arguments.
From this we see that the largest contribution to the $T$
parameter from the couplings $\la_{u,d}$ comes from couplings
such as
\beq[cTterm]
\Delta \lag_{\rm eff} =
\frac{c_T}{16 \pi^2} \left[{\rm tr}(\mathcal H \lambda 
D_\mu \Sigma^\dagger)\right]^2,
\eeq
where $c_T \sim 1$.
This gives 
\beq
\Delta m_W^2 =\Delta m_{W^\pm}^2 - \Delta m_{W_3}^2  \sim \frac{g^2 f^2}{4} (\epsilon_u -\epsilon_d)^2,
\eeq
or
\beq[eq:TUV]
\Delta T_{\rm UV} = \alpha^{-1} \frac{\Delta m_W^2}{m_W^2} \sim \alpha^{-1} (\epsilon_u -\epsilon_d)^2,
\eeq
where the expansion parameters $\ep_{u,d}$ are defined in \Eq{epsilondefn}.
For the values used above, we find
$\De T \sim 0.3$, which is just the right size to get a good
precision electroweak fit (see below).

There is another UV contribution to the $T$ parameter in the strong
sector coming from $U(1)_Y$ loops that is of order
$\De T \sim \pm 1/4\pi$.
This should be regarded as an additional uncertainty on the size of
the $T$ parameter in these models.
This contribution is sufficiently small that it does not affect our
conclusions below.

There are also IR contributions to the $T$ parameter from states below
the scale $\La$.
The largest contribution comes from the light Higgs.
For large $\tan\be$ this is mainly the excitation from $H_u$ and we
have simply
\beq[eq:TIR]
\Delta T_{\rm IR} = -\frac{3}{16 \pi \cos^2 \theta_W}
\ln \frac{m_h^2}{m_{h,{\rm ref}}^2}. 
\eeq
The mass eigenstates
$(A_1^0, H_1^\pm)$ and $(A_2^0, H_2^\pm)$ form approximately
degenerate custodial $SU(2)$ multiplets, and we will neglect their 
contribution to the $T$ parameter.
Note that there is already a large uncertainty in the $T$ parameter
because we only know the order of magnitude of the effective coupling
$c_T$ in \Eq{cTterm}.

\begin{figure}[t]
\begin{center}
\includegraphics[width=10cm]{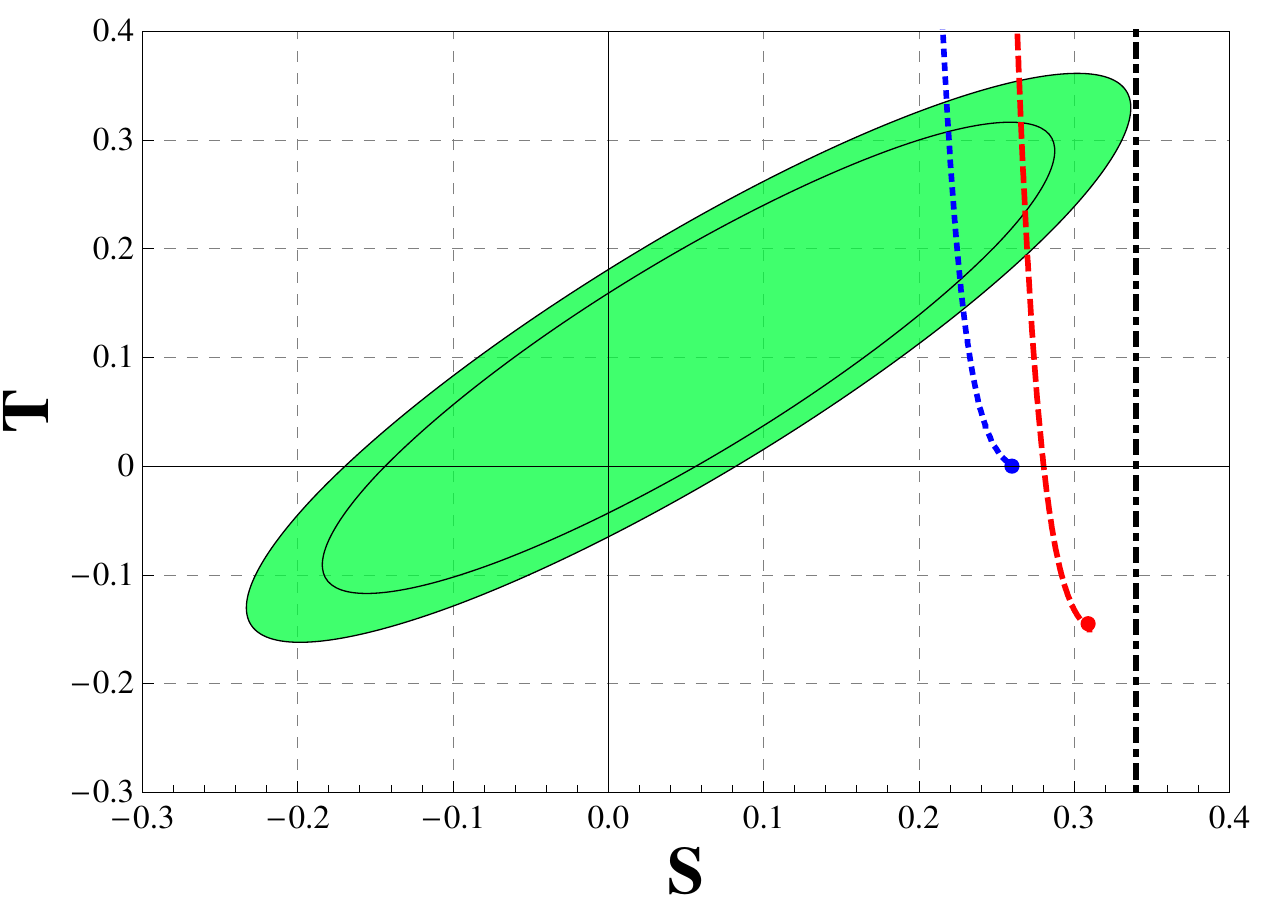}
\caption{Electroweak fit for 
$f = 100\GeV$, $\tan\be = 5$, $B\mu = 0$.
The inner (outer) ellipse is the $95\%$ ($99\%$) confidence level
allowed region for a reference Higgs mass of 120 GeV \cite{Gfitter}.
The dotted blue (dashed red) line corresponds to a light
Higgs mass of 120~(350)~GeV in the model of Section \ref{sec:TCinduced}.
The dot-dashed black line corresponds to the model of Section \ref{sec:CSC}.
As discussed in the text, there are large uncertainties in these
curves; in particular it is plausible that the $S$ parameter is
significantly smaller.
The assumptions that go into these curves are described in the text.}
\label{fig:ST_f100}
\end{center}
\end{figure}

To give some idea of the prospects for a precision electroweak fit,
we plot these estimates in Fig.~\ref{fig:ST_f100}.
We assume that the UV contribution to the $S$ parameter is given
by the QCD value \Eq{SQCDeq} and the UV contribution to the $T$
parameter is given by the \rhs\ of \Eq{eq:TUV}.
We assume that the UV contribution to the $T$ parameter is positive,
as suggested by perturbation theory.
With these assumptions, the plot shows the values of $S$ and $T$
for light Higgs masses of $120\GeV$ and $350\GeV$.
For each Higgs mass there is a line of values corresponding to
different values of custodial
$SU(2)$ violation from the couplings $\la_{u,d}$.
The curves are not entirely in the $T$ direction because
the masses of the heavy PNGB fields depend on these
couplings, so changing these couplings gives a contribution
to $S$ as well as $T$.
For values of the light Higgs mass above $350\GeV$ the expansion
is not under theoretical control because $\la_u$ becomes too large.
The net result is that a positive contribution to the $T$ parameter can
give a good precision electroweak fit under these assumptions,
in the region where the theory is under theoretical control.
There is a large theoretical uncertainty in the predictions for
$S$ and $T$, so the plots cannot be taken too literally,
and our conclusion is that precision electroweak data does not 
strongly constrain these models given our present knowledge.
In fact, the only scenarios we can envision that precision electroweak
can rule out these models is if either the 
$S$ parameter is much larger than expected, or the UV contributions
to the $T$ parameter are negative.
Neither of these is expected.

Finally, we consider $Z \to \bar{b}b$.
the strong sector couples weakly to the elementary
Higgs fields, which have the Yukawa couplings to the top and bottom
quarks.
This means that any correction to $g_{Z \bar{b}b}$ from
the strong sector must be suppressed by $y_t^2$ as well as $\la_{u,d}^2$.
We write the third generation Yukawa couplings as
\beq
\De\scr{L} = Q_L^T \ep \scr{H} y Q_R^c + \hc,
\eeq
where $\scr{H}$ is defined in \Eq{eq:TSpurions} and
\beq[ydefn]
Q_L = \pmatrix{t_L \cr b_L \cr},
\qquad
Q^c_R = \pmatrix{b_R^c \cr t_R^c},
\qquad
y = \pmatrix{y_b & 0 \cr 0 & y_t \cr}.
\eeq
The leading correction to $Z \to \bar{b}b$ comes from 
effective interactions of the form
\beq
\De\scr{L}_{\rm eff} \sim \frac{1}{(4\pi)^4}
Q_L^\dagger \bar \si^\mu Q_L \tr(iD_\mu \Si 
\la y^\dagger y \la^\dagger \Si^\dagger). 
\eeq
This gives a correction
\beq
\frac{\De g_{Z\bar{b}b}}{g_{Z\bar{b}b}}
\sim \frac{y_t^2}{16\pi^2}
\frac{\la_u^2}{16\pi^2}
\sim \frac{y_t^2}{16\pi^2}
\left( \frac{m_h}{4\pi v} \right)^4
\left( \frac{v}{f} \right)^{6}.
\eeq
The standard model agrees with the measured value at the level
of $0.25\%$, which gives the constraint (for $m_h \simeq 120\GeV$)
\beq
v < 5.6 f.
\eeq
This is easily satisfied given the other constraints 
we have already considered above.

\subsection{Collider Phenomenology}
We now discuss the collider phenomenology of this model,
focusing on the LHC.
This theory has SUSY broken at the TeV scale, 
so it has the standard SUSY signals resulting from pair
production of strongly interacting superpartners followed
by cascade decays.
This work focuses on electroweak symmetry breaking, and 
does not prefer any particular pattern of masses for the
MSSM superpartners.

In addition to the standard SUSY signals, this model
extends the MSSM Higgs sector with a custodial $SU(2)$
triplet of PNGBs, which mix with the $CP$ odd Higgs fields
of the MSSM.
The heavy mass eigenstates $A_2^0$ and $H_2^\pm$ are 
dominantly from the strong sector, with $\scr{O}(f/v)$ mixing
with the light MSSM Higgs fields.
The $A_2^0$ can be directly produced via gluon-gluon fusion
through a top quark loop, with a cross section of order 
$f^2/v^2$ times the standard model cross section.
For $m_{A_2^0} = 500\GeV$ this cross section is of order 10~fb
at the LHC.
The $A_2^0$ has potential decay modes
$A_2^0 \to h^0 Z$ and 
$A_2^0 \to A_1^0 h^0$ followed by either
$A_1^0 \to \bar{t}t$ or $Z h^0$.
As we have seen above, we can get a good precision electroweak
fit for large values of the $h^0$ mass, so we can have either
$h^0 \to \bar{b}b$ or $WW$/$ZZ$.
There are many possible final states to investigate, but the
common feature is a high multiplicity of heavy standard 
model particles.

We can also produce heavy hadrons from the strong sector.
These are expected to be at the scale $4\pi f \sim \mbox{TeV}$.
They can be produced via vector boson fusion
(for resonances of spin 0, 1, or 2), or by mixing with
the $W$ and $Z$ (for spin 1).
NDA tells us that the couplings of such a resonance $\rho$ are
\beq
\scr{L}_{\rm eff}
\sim (\d \rho)^2 + \La^2 \rho^2
+ \frac{g}{4\pi} \La^2 \rho W
+ \frac{g^2}{4\pi} \La \rho WW
+ \cdots.
\eeq
This is the same coupling as in traditional technicolor theories,
but with a reduced strong scale $\La$.
The mixing of spin-1 resonances with the $W$ and $Z$ is therefore
of order $g/4\pi$, so we have production of neutral spin-1 resonances
with a cross section suppressed by $g^2 / 16\pi^2$ compared to a
sequential $Z'$ of the same mass.
Production via vector boson fusion is also possible.

These heavy resonances will generally decay to 2-body
final states involving strong particles, \ie
they will pair-produce $A_2^0$ and $H_2^\pm$.
The decays of the $A_2^0$ have been discussed above.
The dominant decays of the heavy charged Higgs fields are
expected to be $H_2^\pm \to W^\pm h^0$ and
$H_2^\pm \to A_1^0 W^\pm$.
The light charged Higgs fields 
can decay via $H_1^+ \to \bar{b}t$ or $W^+ h^0$.
We see that this opens up even more final states
with even higher multiplicity of heavy standard model particles.

It should be clear from this discussion that the phenomenology is
very rich and exciting.
We will leave detailed investigation of LHC signals to
future work.

\section{Strong Electroweak Symmetry Breaking
\label{sec:CSC}}
We now consider another scenario for electroweak symmetry breaking
where there are no elementary Higgs fields below the TeV scale.
The theory at the TeV scale consists of the MSSM \emph{without}
the Higgs fields, plus a strong conformal sector.
SUSY breaking at the TeV scale gives masses to the MSSM superpartners,
and triggers confinement and chiral symmetry breaking in the
strong sector, breaking electroweak symmetry.
Quark and lepton masses arise from interactions between the
strong sector and the quarks and leptons.

As described above, this scenario is very similar to
conformal technicolor.
The main difficulties in constructing a realistic model of 
conformal technicolor are constructing a mechanism to generate
the top quark mass without flavor-changing neutral currents,
and the precision electroweak tests.
The presence of SUSY broken at the TeV scale greatly alleviates
both of these problems, as we will discuss below.
The absence of a light Higgs of course means that the SUSY Higgs
mass problem is absent, which is the main motivation for this
model.

\subsection{Flavor}
We first discuss the origin of the quark and lepton masses.
The strong sector is assumed to contain chiral superfield operators
$\scr{O}_{u,d}$ with the quantum numbers of the MSSM Higgs fields.
These have Yukawa-type couplings with
the quark and leptons superfields that generate
fermion masses.
In any interacting conformal theory the operators $\scr{O}_{u,d}$
have dimension $d > 1$, so the Yukawa interactions are irrelevant
interactions.
(In the model described in Section~\ref{sec:UVmodel}, $d = \frac 32$.)
The general danger in conformal technicolor is that 
$\scr{O}_{u,d}^\dagger \scr{O}_{u,d}$ has dimension $< 4$,
so that there is a relevant singlet operator.
But this operator is not invariant under SUSY, and is therefore
protected from large UV contributions.
This is just a restatement of the well-known fact that scalar mass terms
are forbidden by SUSY, even for fields with $d = 1$.

The Yukawa coupling responsible for the top
quark mass gets strong at a scale $\La_t$ that is quite
low, even for for small values of $d$.
(For $d = \frac 32$, $\La_t \sim 600\TeV$.)
At or below the scale $\La_t$ we need a theory that generates
these interactions without generating additional interactions
that lead to large flavor-changing neutral currents.
These can be generated by exchange of elementary scalars
with the quantum numbers of Higgs doublets \cite{bosonicTC}.
These scalar fields have ordinary Yukawa couplings with quarks
and leptons, and therefore have minimal flavor violation.
(Of course, because the theory is supersymmetric at the TeV scale
we still have to address the SUSY flavor problem associated with
squark and slepton masses and $A$ terms.)
For $\La_t \gg \mbox{TeV}$, getting a sufficiently large
top mass requires that these
scalars have large couplings to the top quark,
the strong sector, or both \cite{CTCflavor}.

An alternative is to have $\La_t \sim \mbox{TeV}$.
This is very natural in the present class of models:
the elementary Higgs scalars can have positive mass-squared
terms of order the TeV scale,
and generate the required couplings at this scale.
The couplings of the elementary Higgs fields to the strong
sector are generally relevant interactions, and so one
must explain why these interactions are important at
the SUSY breaking scale.
This is similar to the problem of explaining why
the $\mu$ term of the MSSM is of order the SUSY breaking
scale, and in the model of Section~\ref{sec:UVmodel} 
we give a solution based on a generalization of the 
Giudice-Masiero mechanism.
If we normalize the Higgs coupling to the strong sector
at the TeV scale like a dimensionless Yukawa coupling
$y_{\rm TC}$, we have
\beq[mtstrong]
m_t \sim y_t y_{\rm TC} v.
\eeq
We see that this requires neither the top quark Yukawa coupling
nor the coupling of the Higgs to the strong sector to be strong.

\subsection{Precision Electroweak Fit}
We now turn to the precision electroweak fit.
Many of the comments made in Section~\ref{sec:PEWinduced}
apply to this case as well, so we will be brief.

We begin with the $S$ parameter.
The strong sector need not have large $N$, and so the
contributions to the $S$ parameter from this sector
is not large to begin with.
In addition, there are good reasons to think that
the UV contribution to the $S$ parameter may be significantly
reduced compared to the QCD value.
This is suggested by recent lattice calculations \cite{Slattice},
and there are theoretical arguments that this occurs in theories
that are conformal above the TeV scale \cite{sundrumhsu}.
The IR contribution to the $S$ parameter is as in technicolor:
\beq
S_{\rm IR} = \frac{1}{12\pi} \ln \frac{\La^2}{m_{h,{\rm ref}}^2},
\eeq
where $\La \sim 4\pi v \sim 3\TeV$.

We now discuss the $T$ parameter.
The couplings of the elementary scalars to the strong sector
that generate quark and lepton masses in general violate custodial
symmetry, and give an additional contribution to the $T$ parameter.
We assume that this contribution is positive (as suggested by
perturbation theory), in which case it can help with the precision
electroweak fit.
There is no limit to how large this contribution can be, since
the couplings of the Higgs fields to the strong sector can
naturally be strong at the TeV scale.
This requires a reduced value for the top quark Yukawa coupling;
see \Eq{mtstrong}.
On the other hand, it is natural for custodial symmetry 
to be an approximate symmetry of this sector, so these
contributions to the $T$ parameter need not be large.

The upshot is that the $T$ parameter is an adjustable parameter in
this model.
This is illustrated in Fig.~\ref{fig:ST_f100}.
Here we have simply assumed the QCD value for the UV contribution
to the $S$ parameter together with an arbitrary positive $T$ contribution.
To get a good precision electroweak fit,
the UV contribution to $S$ must be reduced 
compared to the QCD value, but a factor of 2 is more than
sufficient.
This is clearly within the uncertainties
(see the discussion in Section \ref{sec:PEWinduced}), and we conclude that
precision electroweak is not a strong constraint on these
models given our present state of knowledge.

Finally, we discuss $Z \to \bar{b}b$.
In this model, the strong sector couples directly to the top and bottom
quarks, so the leading correction to $Z \to \bar{b}b$
comes from
effective interactions of the form
\beq
\De\scr{L}_{\rm eff}
\sim \frac{1}{\La^2}
\tr( D_\mu \Si y y^\dagger \Si^\dagger )
Q_L^\dagger  \si^\mu Q_L
+ \hc
\eeq
where $y$ is defined in \Eq{ydefn}.
This gives
\beq
\frac{\De g_{Z\bar{b}b}}{g_{Z\bar{b}b}}
\sim
\frac{y_t^2}{16\pi^2}.
\eeq
The standard model agrees with the measured value at the level
of $0.25\%$, and this contribution is about the same size.
We conclude that this correction is at the level of the
measured precision, but there is no direct conflict.

\subsection{Phenomenology}
Below the scale $\La \sim 4\pi v \sim 3\TeV$ the light states
in this model include the usual MSSM superpartners,
minus the Higgs and Higgsino fields.
The absence of the Higgsino fields simplifies the chargino
and neutralino sectors of the theory.
In particular the lightest neutralino is a mixture of the
Bino and the Wino.
Their mixing is suppressed because the Higgs fields are heavy,
so the only neutralino thermal dark matter candidate is a light
Bino, requiring slepton masses right near the experimental limits
\cite{BinoDM}.
There are of course many other possibilities for dark matter
in supersymmetric theories.

We now turn to the LHC phenomenology of this model.
In addition to the standard SUSY signals, this theory has a 
strong electroweak symmetry breaking sector at the TeV scale.
The minimal model has a strong sector with a $SU(2)_L \times SU(2)_R$
symmetry broken down to the diagonal $SU(2)$.
Non-minimal symmetry breaking patterns with additional
PNGBs are also possible, but are not discussed here.
An important difference from traditional technicolor models is
that the strong sector generally does not have an approximate
parity symmetry that interchanges $SU(2)_L \times SU(2)_R$.
This arises because the technisquarks charged under $SU(2)_L$
and $SU(2)_R$ need not have the same masses.
Since these masses determine the confinement scale, this
breaking of parity is unsuppressed at this scale.
This implies that the resonances that unitarize $WW$ scattering
can generally decay to $WWW$ as well as $WW$.

\section{Conclusions
\label{sec:conclude}}
This work has begun the exploration of models in which SUSY breaking
triggers confinement and chiral symmetry breaking in a strong
sector at the TeV scale.
This is very generic in SUSY gauge
theories with a strong conformal fixed point,
since soft SUSY breaking in the strong sector also breaks conformal
invariance softly.
This generates masses for all scalars in the strong sector, while
fermion masses are generally protected by chiral symmetries.
Since the gauge coupling is strong at all scales, this very plausibly
leads to confinement and chiral symmetry breaking at the SUSY breaking
scale.

We have considered models in which the strong dynamics breaks electroweak
symmetry, in two different limits.
In one limit the strong sector induces large VEVs in elementary Higgs
fields, while in the other the strong dynamics is solely responsible
for electroweak symmetry breaking.
Both of these scenarios can have a good precision electroweak fit
thanks to an adjustable $T$ parameter arising from the elementary
Higgs couplings to the strong sector.
Both have no problems generating the large top quark mass without
additional flavor-changing interactions.
Both scenarios share the usual SUSY flavor problem with the MSSM,
which which may be solved using one of the many mechanisms in the literature.
The important point is that the presence of the strong
dynamics does not give rise to any additional flavor problem.
Unlike the MSSM, gauge coupling unification is no longer a prediction of
the models described here,
since the strong sector affects the evolution
of the $SU(2)_W \times U(1)_Y$ gauge couplings but not
$SU(3)_C$.
Unification can be accommodated with additional matter fields,
which however have no other apparent motivation in this framework.
The phenomenology of these scenarios is rich and deserves further
study.
We also believe that further theoretical investigation of the combination
of SUSY breaking and strong dynamics will be fruitful.

\section*{Acknowledgements}
We thank
R. Contino,
R. Kitano,
T. Okui, 
and 
J. Terning
for discussions.
We also thank J. Serra for participation at early stages of this work,
and M. Schmaltz for emphasizing the importance of
constraints on $Z \to \bar{b}b$.
This work was supported by DOE grant
DE-FG02-91-ER40674.

\appendix{Appendix: Singlet Soft Masses}
We now discuss the effect of a universal soft SUSY breaking mass for the
singlets $S_{ij}$ in the model of Section~\ref{sec:viable}.
The terms in the UV Lagrangian involving $S$ can be written
\beq
\scr{L} = \myint d^4\th\, Z_S S_{ij}^\dagger S_{ij}
+ \left( \myint d^2\th\, \la_{ij} S_{ij} \Psi_i \Psi_j + \hc \right).
\eeq
The universal soft mass can be parameterized by a nonzero
$D$ component for $Z_S$:
\beq
Z_S \sim 1 + D_S \th^4,
\eeq
where $D_S \sim M_{\rm SUSY}^2 \ll \La_*^2$.
We can think of $Z_S$ as a gauge field for a $U(1)_S$ gauge symmetry
under which
\beq
\bal
S_{ij} &\mapsto e^{i\Om} S_{ij},
\\
\la &\mapsto e^{-i\Om} \la,
\\
Z_S &\mapsto e^{i(\Om - \Om^\dagger)} Z_S,
\eal\eeq
where $\Om$ is a chiral superfield gauge transformation parameter.
The fact that $\la \ne 0$ breaks the $U(1)$ gauge symmetry explicitly,
but this breaking is soft in the UV theory.
Another important symmetry is a $U(1)_R$ symmetry with charges
\beq[theRsymmetry]
R(\Psi) = \sfrac 12,
\qquad
R(S) = 1.
\eeq

Now consider this theory below the scale $\La_*$ where the couplings
$\la$ become strong.
The question is then how does the spurion $Z_S$ appear in the low-energy
effective theory?
The low-energy degrees of freedom are the dual techniquarks $\tilde\Psi$
which carry no $U(1)_S$ charge.
The dependence on $Z_S$ is therefore via the $U(1)_S$ gauge
invariant quantities
\beq
\xi &= \frac{\la^\dagger \la}{Z_S},
\\
S_\al &= \bar{D}^2 D_\al \ln Z_S.
\eeq
$\xi$ is proportional to the physically normalized
superpotential coupling strength, while $S_\al$ is the $U(1)_S$
gauge field strength.
These contain SUSY breaking
\beq
\xi &\sim \la^2(1 + \th^4 D_S),
\\
S_\al &\sim \th_\al D_S,
\eeq
and therefore parameterize the SUSY breaking arising from the
$S$ soft mass in the low-energy theory.
For example, the effective theory contains the terms
\beq[DeLeff]
\De\scr{L}_{\rm eff}
\sim \myint d^4\th\, \xi \tilde{\Psi}^\dagger \tilde{\Psi}.
\eeq
This gives a universal soft mass for the dual techniquarks.
Since the operator $\tilde{\Psi}^\dagger \tilde{\Psi}$
has dimension $> 2$, this operator becomes important at a scale
parametrically below $M_{\rm SUSY}$.

There can be terms in the effective Lagrangian
proportional to strong operators
that are not singlets,
which are not required to be irrelevant operators.
These all involve the spurion $S_\al$ since $\xi$ is a singlet under all
symmetries.
It is easily checked that there are no allowed $F$ terms involving
$S_\al$ allowed by $U(1)_R$ symmetry.
We can systematically enumerate all $D$ terms involving $S_\al$.
An example is
\beq
\myint d^4\th\, S^\al \scr{O}_\al
= D_S
\times \bar{D}^2 D^\al \scr{O}_\al |_{\th = 0}.
\eeq
Unitarity requires $\dim(\scr{O}_\al) > \frac 32$, so the
operator on the \rhs must have
dimension $> \frac 32 + \frac 32 = 3$.
Since the theory is strongly coupled, we expect this inequality
to be violated by $\scr{O}(1)$.
Matching at the scale $\La_*$ and running down, we see that
dimensionless strength of this SUSY breaking is 
\beq
\de \ll \left( \frac{D_S}{\La_*^2} \right)^2
\left(\frac{E}{\La_*} \right)^{-2}.
\eeq
This 
gets strong at a scale
\beq
E \ll \frac{M_{\rm SUSY}^2}{\La_*} \ll M_{\rm SUSY}.
\eeq
Similarly, we have
\beq
\myint d^4\th\, D^\al S_\al \scr{O} = D_S \times
D^2 \bar{D}^2 \scr{O} |_{\th = 0}
&\quad\Rightarrow\quad
\mbox{dim} > 3,
\eql{op1}
\\
\myint d^4\th\, S^\al S_\al \scr{O} =
D_S^2 \times \bar{D}^2 \scr{O}  |_{\th = 0}
&\quad\Rightarrow\quad
\mbox{dim} > 2,
\\
\myint d^4\th\, S^\al (S^\dagger)^{\dot\al}
\scr{O}_{\al\dot\al} =
D_S^2 \times D^\al \bar{D}^{\dot\al} \scr{O}_{\al \dot\al}  |_{\th = 0}
&\quad\Rightarrow\quad
\mbox{dim} > 4,
\\
\myint d^4\th\, S^\al S_\al (S^\dagger)^{\dot\al} \scr{O}_{\dot\al} =
D_S^3 \times \bar{D}^{\dot\al} \scr{O}_{\dot\al}  |_{\th = 0}
&\quad\Rightarrow\quad
\mbox{dim} > 2,
\eql{op4}
\\
\myint d^4\th\, |S^\al S_\al|^2 \scr{O} =
D_S^4 \scr{O}  |_{\th = 0}
&\quad\Rightarrow\quad
\mbox{dim} > 2.
\eql{op5}
\eeq
In \Eqs{op1}--\eq{op4} we used the unitarity constraint on the
dimension of operators, while in \Eq{op5} we used the fact that
$\scr{O}$ is a $R = 0$ operator, and therefore the operator
$\myint d^4\th\, \scr{O}$ is an allowed term in the Lagrangian,
so $\scr{O}$ must have dimension $> 2$.
All of these terms become important at scales parametrically below
$M_{\rm SUSY}$.
Terms with additional derivatives are even more suppressed.
We conclude that all possible SUSY breaking terms in the
low-energy theory are suppressed compared to $M_{\rm SUSY}$.

\end{document}